\documentclass[12pt,preprint]{aastex} 
\shorttitle{} 
\shortauthors{} 

\newcommand\lsim{\mathrel{\rlap{\lower4pt\hbox{\hskip1pt$\sim$}}
        \raise1pt\hbox{$<$}}}
\newcommand\gsim{\mathrel{\rlap{\lower4pt\hbox{\hskip1pt$\sim$}}
        \raise1pt\hbox{$>$}}}
 
\begin{document} 
  
\title{Can Supermassive Black Holes Form in Metal-Enriched High-Redshift Protogalaxies ?} 
\author{K. Omukai \altaffilmark{1},  
R. Schneider \altaffilmark{2} and  
Z. Haiman \altaffilmark{3}} 
\altaffiltext{1}{National Astronomical Observatory of Japan,  Mitaka, Tokyo 181-8588, Japan; omukai@th.nao.ac.jp} 
\altaffiltext{2}{INAF--Osservatorio Astrofisico di Arcetri,  Largo E. Fermi 5, 50125 Florence, Italy; raffa@arcetri.astro.it} 
\altaffiltext{3}{Department of Astronomy, Columbia University,  550 West 120th Street, New York, NY 10027, USA; zoltan@astro.columbia.edu} 
 
\begin{abstract} 
Primordial gas in protogalactic dark matter (DM) halos with virial
temperatures $T_{\rm vir}\gsim 10^4$K begins to cool and condense via
atomic hydrogen.  Provided this gas is irradiated by a strong
ultraviolet (UV) flux and remains free of ${\rm H_2}$ and other
molecules, it has been proposed that the halo with $T_{\rm vir}\sim 10^4$K
may avoid fragmentation, and
lead to the rapid formation of a supermassive black hole (SMBH) as
massive as $M\approx 10^5-10^6~{\rm M_\odot}$.  This ``head--start''
would help explain the presence of SMBHs with inferred masses of
several $\times 10^9{\rm M_\odot}$, powering the bright quasars
discovered in the Sloan Digital Sky Survey at redshift $z\gsim 6$.
However, high--redshift DM halos with $T_{\rm vir}\sim 10^4$K are
likely already enriched with at least trace amounts of metals and dust
produced by prior star--formation in their progenitors.  Here we study
the thermal and chemical evolution of low--metallicity gas exposed to
extremely strong UV radiation fields. Our results, obtained in
one--zone models, suggest that gas fragmentation is inevitable above a
critical metallicity, whose value is between $Z_{\rm cr} \approx 3
\times 10^{-4} Z_{\odot}$ (in the absence of dust) and as low as
$Z_{\rm cr} \approx 5 \times 10^{-6} Z_{\odot}$ (with a dust-to-gas mass ratio
of about $0.01 Z/Z_{\odot}$).  We propose that when the metallicity exceeds these
critical values, dense clusters of low--mass stars may form at the
halo nucleus. 
Relatively massive stars in such a cluster can then rapidly coalesce
into a single more massive object, which may produce an intermediate--mass BH
remnant with a mass up to $M\lsim 10^2-10^3~{\rm M_\odot}$.
\end{abstract}  
 
\keywords{cosmology: theory --- galaxies: formation --- stars: formation} 
 
\section{Introduction}
\label{sec:intro}

The discovery of bright quasars at redshifts $z\gsim 6$ in the Sloan
Digital Sky Survey (SDSS) implies that BHs as massive as several $\times
10^9{\rm M_\odot}$ were already assembled when the age of the universe
was less than $\approx 1$ Gyr (see the recent review by Fan 2006).
The BH masses are inferred from the quasars' luminosities, assuming
these sources shine near their Eddington limit.  Strong gravitational
lensing or beaming could, in principle, mean that the inferred BH
masses are overestimated; however, there is no obvious sign of either
effect in the images and spectra of these quasars (Willott et
al. 2003; Richards et al. 2004).  Indeed, their relatively ``normal''
line--to--continuum ratio, consistent with those in lower--redshift
quasars, makes it unlikely that the apparent flux of these sources was
significantly boosted by beaming (Haiman \& Cen 2002). Likewise, the
lack of a second detectable image on {\it Hubble Space Telescope}
images (Richards et al. 2004) essentially rules out the hypothesis
that most of the sources experienced strong magnification by lensing
(Comerford et al. 2002; Keeton et al. 2005).

Relatively little time is available for the growth of
several$\times10^9~{\rm M_\odot}$ SMBHs prior to $z\sim 6$, and their
seed BHs must be present as early as $z\sim 10$ (e.g. Haiman \& Loeb
2001). As the SMBHs grow from high--redshift seed BHs by accretion,
they are expected to encounter frequent mergers.  A coalescing BH
binary experiences a strong recoil due to gravitational waves (GWs)
emitted during the final stages of their merger. The typical recoil
speed is expected to be $v_{\rm recoil}\gsim 100~{\rm km~s^{-1}}$ (and
may be as large as $4,000~{\rm km~s^{-1}}$ for special BH spin
configurations; see, e.g.  Campanelli et al.  2007 and references
therein), significantly exceeding the escape velocity ($\lsim 10~{\rm
km~s^{-1}}$) from typical DM halos that exist at $z\sim 10$. As a
result, SMBHs are often ejected from their host halos at high
redshift. The repeated loss of the growing seeds makes it especially
challenging to account for the several$\times10^9~{\rm M_\odot}$ SMBHs
at $z\gsim 6$ without at least a brief phase of super--Eddington
accretion, or some equivalent ``head--start'' (Haiman 2004; Yoo \&
Miralda-Escud\'{e} 2005; Shapiro 2005; Volonteri \& Rees 2006).

There have been several recent proposals that such a ``head--start''
may occur in metal--free gas in high--redshift DM halos with virial
temperatures exceeding $T_{\rm vir}\gsim 10^4$K, leading to the rapid
formation of SMBHs with a mass of $M\approx 10^5-10^6~{\rm M_\odot}$.
As primordial gas falls into these halos, it initially cools via the
emission of hydrogen Ly$\alpha$ photons. Provided the gas is free of
${\rm H_2}$ molecules, its temperature will remain near $T_{\rm
vir}\sim 10^4$K.  Bromm \& Loeb (2003, hereafter BL03) performed
hydrodynamical simulations of a metal-- and ${\rm H_2}$--free halo,
with a mass of $\sim 10^{8}M_{\sun}$ collapsing at $z \sim 10$,
corresponding to a 2$\sigma$ Gaussian overdensity and to $T_{\rm
vir}\sim 10^{4}$~K.  Under these conditions, which may apply to some
dwarf galaxies collapsing close to the epoch of reionization, the
primordial gas is marginally able to collapse and remains nearly
isothermal.  BL03 found that during the evolution, fragmentation of
the gas cloud is very inefficient, leading at most to binary formation
even with some degree of rotation. Thus, a super--massive star is
expected to form, and evolve into a SMBH with a mass as high as
$M\approx 10^5-10^6~{\rm M_\odot}$. 
Oh \& Haiman (2002) and Lodato \& Natarajan (2006) have also showed
that if H$_2$ formation is inhibited, a primordial-gas disk is stable
to fragmentation and a single massive object is formed in accordance
with BL03's conclusion.
Volonteri \& Rees (2005) arrived
at similar conclusions, by considering Bondi accretion onto a stellar
seed BH, which can significantly exceed the Eddington rate at the gas
density and temperature in a similar halo.  
Finally, Begelman et
al. (2006) and Spaans \& Silk (2006) proposed different mechanisms to
form similarly massive BHs by the direct collapse of primordial,
atomic gas.  For reference, we note that the total (DM+gas) mass of
halos with $T_{\rm vir}=(1-5)\times 10^4$K at $z=10$ is $M_{\rm
tot}\approx 10^{8-9}~{\rm M_\odot}$, so that such SMBHs would
represent $\approx 0.2-20\%$ of the gas mass in these halos.  We also
note that in the WMAP5 cosmology, the age of the universe at $z=10$
and $z=6.5$ is $\sim 0.5$Gyr and $\sim 0.9$ Gyr, respectively. At the
e--folding time--scale of $4\times 10^7$ years (assuming Eddington
accretion, and a radiative efficiency of 10\%; see, e.g., Haiman \&
Loeb 2001), 
a seed BH of $M\approx 10^5{\rm M_\odot}$ at $z\sim 10$ {\it could} 
easily grow to a super massive BH of $M\approx 2\times 10^9 {\rm M_\odot}$ 
at $z\sim 6.5$, if fed uninterruptedly.

{\em A crucial assumption in all of the above proposals is that $H_2$
molecules cannot form as the gas cools and condenses in the DM halo}.
This assumption can be justified in the presence of a sufficiently
strong far ultraviolet (FUV) radiation, so that molecular hydrogen (or
the intermediary ${\rm H^-}$ necessary to form ${\rm H_2}$) is
photodissociated.
The relevant criterion is that the photodissociation timescale is
shorter than the ${\rm H_2}$--formation timescale; since generically,
$t_{\rm diss}\propto J$ and $t_{\rm form}\propto \rho$, the condition
$t_{\rm diss}=t_{\rm form}$ yields a critical flux $J \propto \rho$.
In DM halos with $T_{\rm vir}\lsim 10^4$K, whose gas can not cool in
the absence of ${\rm H_2}$, the densities remain low and ${\rm H_2}$
can be dissociated even when background flux is as low as
$J_{21}^{-}\sim 10^{-2}$ (e.g. Haiman, Rees \& Loeb 1997; Mesinger et
al. 2006; here $J_{21}^{-}$ is the flux just below $13.6$eV, in the
usual units of $10^{-21} {\rm erg~cm^{-2}~sr^{-1}~s^{-1}~Hz^{-1}}$).
However, if a gas cloud is massive enough and has a virial temperature
higher than $\approx 8000$~K, it is able to cool and start its
collapse via atomic hydrogen Ly-$\alpha$ cooling.  Even if the FUV
field is initially above the critical value, molecular hydrogen can
form, and dominate the gas cooling at a later stage during the
collapse (Oh \& Haiman 2002); the ${\rm H_2}$--formation rate is
furthermore strongly boosted by the large out--of--equilibrium
abundance of free electrons in the collisionally ionized gas in these
halos (Shapiro \& Kang 1987; Susa et al. 1998; Oh \& Haiman 2002).
The critical flux required to keep the gas ${\rm H_2}$--free as it
collapses by several orders of magnitude therefore increases
significantly; for halos with $T_{\rm vir}\sim 10^4$K the value has
been found to be $J_{21}^{-}\approx 10^{3}-10^{5}$, depending on the
assumed spectral shape (Omukai 2001, hereafter O2001; BL03).  
In halos exposed to such
extremely intense UV fields, the gas cloud is still able to collapse
only via atomic hydrogen line cooling, namely Ly $\alpha$ and H$^{-}$
free--bound (f-b) emission (O2001).

One possible source of such an intense UV field is the intergalactic
UV background just before the epoch of cosmic reionization (BL03).
The ionizing photon flux $J_{21}^{+}$ can be evaluated from the number
density of hydrogen atoms in the intergalactic medium (IGM) and the
average number of photons needed to ionize a hydrogen atom
$N_{\gamma}$, which, in general, is $>1$, owing to recombinations in
an inhomogeneous IGM.  Using the escape fraction of ionizing radiation
$f_{\rm esc}$, the flux $J_{21}^{-}$ just below the Lyman limit is
given by
\begin{equation}
J_{21}^{-}=
\frac{J_{21}^{+}}{f_{\rm esc}} 
\simeq \frac{1}{f_{\rm esc}} \frac{hc}{4 \pi} \frac{N_{\gamma} Y_{\rm H} \rho_{\rm b}}{m_{\rm H}}
\simeq 4 \times 10^3
\left(\frac{N_\gamma}{10}\right) 
\left(\frac{f_{\rm esc}}{0.01}\right)^{-1}
\left(\frac{1+z}{11}\right)^3,
\label{eq:Jreion} 
\end{equation} 
where $Y_{\rm H}=0.76$ is the mass fraction of hydrogen, $m_{\rm H}$ is the
proton mass, and
$\rho_{\rm b}$ is the baryon density (assumed here to correspond to
$\Omega_b h^2=0.023$; Dunkley et al. 2008).  Equation~\ref{eq:Jreion}
shows that $J_{21}^{-}$ can approach the critical value at $z\gsim
10$, provided $f_{\rm esc}$ is small, and $N_\gamma$ is large; $f_{\rm
esc}/N_\gamma \lsim 10^{-3}$.  Although the value of $f_{\rm esc}$ is
quite uncertain, in low--mass minihalos, the expectation is $f_{\rm
esc}\approx 1$, as these halos are easily self--ionized, and most of
their ionizing radiation escapes into the intergalactic medium
(Kitayama et al. 2004; Whalen, Abel \& Norman 2004).  
Observations of nearby star forming
galaxies indicate lower values $f_{\rm esc} < 0.1$ (Leitherer et
al. 1995, Inoue et al. 2005); nevertheless, it appears unlikely that
the mean cosmic background will reach the critical values. A few halos
that have close and bright neighbors may still see a sufficiently
increased flux (Dijkstra et al. 2008).  Alternatively, the critical
value could be established by sources internal to the halo, e.g. by a
vigorous phase of starburst (Omukai \& Yoshii 2003) or by the
accreting seed BH itself (Volonteri \& Rees 2005).

So far, the process of supermassive star (and SMBH) formation in dwarf
galaxies in the presence of a strong UV background has been
investigated only under the hypothesis that the gas is metal--free.
However, the $T_{\rm vir}\sim 10^4$K, $M\sim 10^8~{\rm M_\odot}$ halos
forming close to the epoch of reionization are built from lower--mass
progenitors that had collapsed earlier.  Many, and perhaps all of
these halos should therefore be enriched with at least some trace
amount of metals.  Furthermore, is it unlikely that the strong
critical FUV flux could be generated and maintained without efficient
star formation at higher redshifts (if $J_{21}\gsim 10^3$ was produced
by accreting BHs, this would significantly overpredict the
present--day soft X--ray background; Dijkstra et al. 2004).  It is
well--known that adding metals and dust into a primordial gas, even at
trace amounts as low as $Z\sim 10^{-6} {\rm Z_\odot}$, can
significantly affect its cooling properties (Schneider et al. 2003; 
Omukai et al. 2005; Schneider et al. 2006).  Two independent
hydrodynamic simulations (Tsuribe \& Omukai 2006; Clark, Glover, \&
Klessen 2008) recently studied the fragmentation of metal--enriched
collapsing protogalactic clouds, in the absence of an external FUV
field, and found efficient fragmentation for $Z\gsim 10^{-6} {\rm
Z_\odot}$.

In the present paper, our goal is to answer the following question:
{\it can cooling and fragmentation be avoided in metal--enriched
$T_{\rm vir}\gsim 10^4$K halos, irradiated by a strong FUV flux?}  If
so, this would suggest that supermassive black holes may form, similar
to the metal--free case, in the more likely case of metal--enriched
high--redshift protogalaxies.  To investigate this possibility, we
here study the thermal and chemical evolution of low--metallicity gas,
exposed to extremely strong UV radiation fields.  We will evaluate the
critical metallicity, above which fragmentation becomes unavoidable in
the presence of a strong FUV flux.

In \S~\ref{sec:model}, we describe our one--zone modeling procedure.
Our results are presented and discussed in \S~\ref{sec:results}, first
for the metal--free (\S~\ref{sec:metal-free}), and then for the
metal--enriched case (\S~\ref{sec:metal}).  The fragmentation and
subsequent evolution of the metal--enriched clouds are then discussed
in \S~\ref{sec:fragmentation} and \ref{sec:corecollapse},
respectively.  In \S~\ref{sec:conclusions}, we summarize our results
and offer our conclusions.

\section{Model}
\label{sec:model}

\subsection{Basics}

We use the one--zone model described in Omukai (2001) to follow the
gravitational collapse of gas clouds. The model includes a detailed
description of gas--phase chemistry and radiative processes, and 
the effect of dark matter on the dynamics in a simplified fashion.
In addition, in the present version of the model we have implemented 
the contribution of metal lines and dust to gas cooling.

In what follows, all physical quantities are evaluated at the center of the cloud.
The gas density increases as 
\begin{equation}
\frac{d \rho_{\rm gas}}{dt}=\frac{\rho_{\rm gas}}{t_{\rm col}}.
\end{equation}
where the collapse timescale, $t_{\rm col}$, is taken to be equal to the free-fall time,
\begin{equation}
t_{\rm col}= t_{\rm ff} \equiv \sqrt{\frac{3 \pi}{32 G \rho}},
\label{eq:tff}
\end{equation}
and $\rho$ is the sum of the gas and dark matter density.
The dark matter density follows the evolution of a top--hat overdensity,
\begin{equation}
\rho_{\rm DM}=\frac{9 \pi^2}{2} \left(\frac{1+z_{\rm ta}}
{1-{\rm cos}\theta}  \right)^{3} \Omega_{\rm DM} \rho_{\rm crit}
\end{equation}
with
\begin{equation}
1+z=(1+z_{\rm ta}) 
\left( \frac{\theta -{\rm sin}\theta}{\pi} \right)^{-2/3}
\label{eq:zta}
\end{equation}
(e.g., Chapter 8.2 of Padmanabhan 1993),
where the turn-around and the virialization correspond to 
$\theta = \pi$ and $2\pi$, respectively.
Although, strictly speaking, this is correct only in the Einstein-de 
Sitter universe ($\Omega_0=1$), it does not cause a significant 
error in the high-$z$ universe ($z \ga 10$) we consider. 

The initial epoch of calculation is taken at the turn--around 
at redshift $z_{\rm ta}=17$.
From equation \ref{eq:zta}, the
virialization and turn-around redshifts have the relation 
$1+z_{\rm vir}=2^{-2/3}(1+z_{\rm ta})$; 
thus $z_{\rm vir}\simeq 10$.
In our calculation, the dark matter density is kept constant 
after reaching its virialization value $8\rho_{\rm DM}(z_{\rm ta})$.
The initial values of
the gas number density, temperature, ionization degree, and H$_2$
fraction have been assumed to be $n_{\rm H}=4.5 \times 10^{-3} {\rm
cm^{-3}}$, $T=21$~K, $y(e)=3.7 \times 10^{-4}$ and $y({\rm H_{2}})=2
\times 10^{-6}$, respectively, to reflect conditions at the
turn--around at $z_{\rm ta}=17$.
Some runs with initial temperature ten times higher (210K) are 
also performed to confirm independence of our main results from the 
initial temperature.
The cosmological parameters are $\Omega_{\rm DM}=0.24$, 
$\Omega_{\rm b}=0.04$, and $h=0.7$.

Our calculation does not include the virialization shock.
Owing to fast cooling by Ly$\alpha$ emission, the central region whose
evolution we intend to follow does not experience the virialization shock 
in the spherically symmetric case (Birnboim \& Dekel 2003).
In more realistic calculations, the outer regions can experience shocks
and the temperature and electron fraction become higher than in our case.
In addition, recent numerical calculations (e.g. Kere$\check{\rm s}$ et
al. 2005) show that low-mass galaxies, especially at high-redshifts,
obtain their gas through accretion predominantly along the large-scale
filaments.  Three-dimensional effects such as asymmetric accretion
might affect the evolution at low densities.
However, since we are considering halos with $T_{\rm vir}\simeq 10^{4}$K,
which can marginally collapse by  Ly$\alpha$ cooling, the shock is 
not strong: the temperature increase is modest and 
the electron fraction reaches at most $\la 10^{-2}$ (see Figures 5a and 5c in BL03).
This additional electrons alter the early evolution for the $J=0$ case. 
However, in the irradiated clouds, where H$_2$ formation is suppressed,
during the collapse by the Ly$\alpha$ cooling
recombination proceeds until the free electron fraction reaches 
$x_{e}\simeq 1.2\times 10^{-3} n_{\rm H}^{-1/2}$, 
the value set by the balance between the recombination and the collapse time
$t_{\rm rec} \sim t_{\rm col}$ at 8000K. 
Thus, our results for molecule formation and cooling are hardly affected.

We adopt $t_{\rm ff}$ as the collapse time scale just 
because it has been widely used in other studies (e.g., Palla et al. 1983).
Note that the free--fall time (\ref{eq:tff}) is the time for density of 
an initially static cloud to reach infinity, while 
the dynamical timescale $t_{\rm col}=\rho/(d \rho/dt)$ 
in the free--fall collapse is 
\begin{equation}
t_{\rm col, ff}=\frac{1}{\sqrt{24 \pi G \rho}}
\end{equation}
in the limit where the density has become sufficiently 
larger than the initial value.
Thus the rate we adopted (\ref{eq:tff}) is $3 \pi/2=4.7$ times 
slower collapse than the genuine free--fall one. 
In fact, pressure gradients oppose gravity and the 
collapse becomes slower than the free--fall one within a factor of a few 
(e.g., Foster \& Chevalier 1993). 
Adoption of $t_{\rm ff}$ as the e--folding time for 
density increase mimics the pressure effect.
The assumption of nearly free-fall collapse is invalidated, and the
collapse is slowed down, once the cloud becomes optically thick to
continuum radiation.  However, our result on the thermal evolution is
not altered: with little radiative cooling, the temperature is now
determined by the adiabatic compression and the chemical cooling by
dissociation and ionization, both of which are independent of the
collapse timescale.  Moreover, the evolution after the cloud becomes
optically thick is not relevant to our argument on fragmentation,
which occurs at much lower density, in the optically thin regime.

The overall size of the collapsing gas cloud (or of the roughly
uniform density central region) determines its optical depth, and is
therefore important for its thermal evolution.  
Here we assume the size equals the Jeans length,
\begin{equation}
\lambda_{\rm J}=\sqrt{\frac{\pi k T_{\rm gas}}{G \rho_{\rm gas} \mu m_{\rm H}}},
\end{equation}
where $T_{\rm gas}$ is the gas temperature, $\mu$ is the mean molecular weight.
Similarly, its mass is given by the Jeans mass 
\begin{equation}
M_{\rm J}=\rho_{\rm gas} \lambda_{\rm J}^{3}.
\end{equation}
Specifically, we assume that the radius of the cloud is $R_{\rm c}=\lambda_{J}/2$
and the optical depth is
\begin{equation}
\tau_{\nu}=\kappa_{\nu} R_{\rm c}= \kappa_{\nu}\left(\frac{\lambda_{\rm J}}{2}\right).
\end{equation}
In addition to dust absorption (see \S~\ref{sec:dust}), 
we include the following primordial-gas processes as sources of 
the opacity $\kappa_{\nu}$ (Table 1 of O2001): 
the bound-free absorption of H, He, H$^{-}$, H$_{2}^{+}$, 
free-free absorption of H$^{-}$, H, collision-induced absorption of 
H$_{2}$-H$_{2}$ and H$_{2}$-He, Rayleigh scattering of H, and 
Thomson scattering of electrons.

The temperature evolution is followed by solving the energy equation:
\begin{equation}
\frac{d e}{dt}=-p \frac{d}{dt} \left(\frac{1}{\rho_{\rm gas}}\right)
- \frac{\Lambda_{\rm net}}{\rho_{\rm gas}},
\label{eq:energy}
\end{equation}
where $e$ is the internal energy per unit mass
\begin{equation}
e=\frac{1}{\gamma_{\rm ad}-1} \frac{kT_{\rm gas}}{\mu m_{\rm H}},
\end{equation}
$p$ is the pressure, $\gamma_{\rm ad}$ is the adiabatic exponent, and
$\Lambda_{\rm net}$ is the net cooling rate per unit volume.  In
addition to cooling and heating processes for the primordial gas,
which include continuum emission, as well as emission by H and H$_2$
lines, and chemical heating/cooling, the net cooling rate includes
emission by C and O fine--structure lines $\Lambda_{\rm metal}$, by
dust grains $\Lambda_{\rm gr}$, and heating by photoelectric emission
of dust grains $\Gamma_{\rm pe}$.  Cooling by fine--structure lines of
[CII] and [OI] is included as in Omukai (2000).  Dust processes are
described below in \S~\ref{sec:dust}.

Primordial--gas chemical reactions are solved for the nine species of
H, H$_{2}$, $e$, H$^{+}$, H$_{2}^{+}$, H$^{-}$, He, He$^{+}$, and
He$^{++}$. We do not explicitly include the chemical reactions
involving metals.  Instead, all the carbon and oxygen is assumed to be
in the form of CII and OI, respectively. 
Having a lower ionization energy (11.26~eV) than hydrogen, 
carbon remains in the form of CII in the atomic medium owing to 
photoionization by the background radiation.
We maintained this assumption even in $J=0$ runs, although carbon 
is expected to recombine and become neutral in these cases.
The cooling rates by CII and CI fine--structure lines are within a 
factor of $\simeq 2$ difference for $T \ga 30$K, and
therefore this assumption does not significantly affect the results. 
On the other hand, the ionization potential of oxygen (13.61~eV) is
very similar to that of hydrogen (13.60~eV) and the charge exchange 
reaction
\begin{equation}
{\rm  O^{+} + H \leftrightarrow H^{+} + O},
\end{equation}
keeps its ionization degree equal to that of hydrogen.
In fact, the coefficient of the rightward reaction being 
$6.8 \times 10^{-10}{\rm cm^{3}/sec}$, these reactions reach equilibrium 
only in $\sim 50n_{\rm H}^{-1}$yr. 

In a cold ($\la$ a few 100K) and dense ($\ga 10^{3-4}{\rm cm^{-3}}$) 
environment, molecular coolants such as CO and H$_2$O
may become important (Omukai et al. 2005).
Since we are interested here in 
metal effects on warm ($\ga$ a few 1000K) atomic clouds, 
we neglect the contribution to cooling
of metals in molecules. 
This simplification does not affect the early evolution
of gas clouds, when the effects of metals induce a deviation from the 
primordial evolutionary track at several 1000K.
It is true that it may alter the predicted thermal behavior 
at later stages, when the gas has cooled significantly ($\la$ 1000K). 
However, even in such cold environments, the error in the temperature
caused by neglect of metal molecular coolants is very small 
(see Figure 10 of Omukai et al. 2005) and  
the thermal evolution is well reproduced when only dust processes 
and fine--structure line cooling of C and 
O are considered. 
\subsection{Dust Processes}
\label{sec:dust}

Dust in the local interstellar medium (ISM) originates mainly from the
asymptotic giant--branch (AGB) stars, whose age is $\ga$ 1 Gyr, longer
than the Hubble time at $z \ga 6$. At higher redshifts, supernovae
(SNe) are considered to be the major dust factories.  Indeed, the
observed extinction law of high--$z$ quasars and gamma--ray bursts can
be well reproduced by this scenario (Maiolino et al. 2004, Stratta et
al. 2007). Dust grains produced in SN ejecta are more effective in
cooling and H$_2$ formation because of their smaller size and larger
area per unit mass (Schneider et al. 2006).  However, their
composition and size distribution are still affected by many
uncertainties, such as the degree of mixing in the ejecta and the
efficiency of grain condensation and their destruction by the
reverse shock (Nozawa et al. 2007, Bianchi \& Schneider 2007).

To be conservative, in this work the properties of dust, such as grain
composition and size distribution, are assumed to be similar to those
in the solar neighborhood and its amount is reduced in proportion to
the assumed metallicity of the gas clouds.  Specifically, we adopt the
dust opacity model developed by Semenov et al. (2003).  This model
partly follows the scheme proposed by Pollack et al. (1994), which was
used in Omukai et al. (2005), assuming the same dust composition, size
distribution and evaporation temperatures, but uses a new set of dust
optical constants. Overall, the opacity curves of the two models are
in good agreement, the largest difference being at most a factor of
two (see Semenov et al. 2003 for a thorough discussion).  The main
dust constituents include amorphous pyroxene ([Fe, Mg]SiO$_3$),
olivine ([Fe, Mg]$_2$SiO$_4$), volatile and refractory organics,
amorphous water ice, troilite (FeS) and iron. The grains are assumed
to follow a size distribution modified from that 
by Mathis, Rumpl, \& Nordsieck (1977) with the inclusion of large
(0.5 - 5)$\mu$m grains.

At each density and gas temperature, the dust is assumed to be in
thermal equilibrium, and its temperature $T_{\rm gr}$, which is
followed separately from the gas temperature, is determined by the
energy balance equation
\begin{equation}
 4\pi \int \kappa_{{\rm a}, \nu} B_{\nu}(T_{\rm gr}) d\nu
= \Lambda_{\rm gas \rightarrow dust}+
 4\pi \int \kappa_{{\rm a}, \nu} J_{\nu}^{\rm in} d\nu.
\end{equation} 
Here $\Lambda_{\rm gas \rightarrow dust}$ is the energy transfer rate
per unit mass from gas to dust due to gas--dust collisions, which we take
from Hollenbach \& McKee (1979), $\kappa_{{\rm a}, \nu}$ is the
absorption opacity of dust, and $J_{\nu}^{\rm in}$ is the mean
intensity of the radiation field inside the cloud.  Note that
$\Lambda_{\rm gas \rightarrow dust}$ also represents the net cooling
rate of the gas, caused by the presence of dust grains at temperature
$T_{\rm gr}$.
We model the external radiation field assuming a diluted thermal
spectrum (i.e. a blackbody spectrum, scaled by an overall constant
representing a mean geometrical dilution).  Its shape is then fully
described by only two free parameters, $J_{21}$, the mean intensity at
the Lyman limit ($\nu_{\rm H}$) and $T_{\ast}$, the color temperature,
\begin{equation}
J_{\nu}^{\rm ex}=J_{21} 10^{-21} [B_{\nu}(T_{\ast})/B_{\nu_{\rm H}}(T_{\ast})] \,\,  {\rm erg~cm^{-2}sr^{-1}s^{-1}Hz^{-1}}.
\end{equation}
In the following, we will consider two possible values for the
radiation color temperature, $T_{\ast}=10^{4}$~K and $10^{5}$~K,
representing ``standard'' Population II stars and very massive
Population III stars, respectively.  Given the mean intensity of the
external radiation field $J_{\nu}^{\rm ex}$, the field inside the gas
cloud is obtained as (see O2001),
\begin{equation}
J_{\nu}^{\rm in}=\frac{J_{\nu}^{\rm ex} + \xi_{\nu} x_{\nu} S_{{\rm a}, \nu}}
{1 + \xi_{\nu} x_{\nu}},
\end{equation}
where $1-\xi_{\nu}$ is the single--scattering albedo,  
$S_{{\rm a}, \nu}$ is the source function,
\begin{equation}
x_{\nu}={\rm max}[\tau_{\nu}^2,\tau_{\nu}],
\end{equation} 
and the optical depth $\tau_{\nu}$ includes both dust and gas
opacities.

The gas is heated by photoelectrons ejected from dust grains 
after absorption of FUV photons.
Following Bakes and Tielens (1994), we compute the photoelectric
heating rate as,
\begin{eqnarray}
\Gamma_{\rm pe}^{\rm (net)} &=& \Gamma_{\rm pe}-\Lambda_{\rm pe} \\
&=& [10^{-24} \epsilon G_{0} n_{\rm H}
-4.65 \times 10^{-30} T^{0.94} (G_{0} T^{1/2}/n(e))^{\beta} n(e) n_{\rm H}]
Z/Z_{\sun}
\label{eq:peheat}
\end{eqnarray}
where
\begin{equation}
\epsilon=\frac{4.9 \times 10^{-2}}
{[1 + 4 \times 10^{-3} (G_{0} T^{1/2}/n(e))^{0.73}]}
+\frac{3.7 \times 10^{-2} (T/10^{-4})^{0.7}}
{[1 + 2 \times 10^{-4}(G_{0} T^{1/2}/n(e))]}, 
\end{equation}
and $\beta=0.735/T^{0.068}$.
The Habing parameter $G_{0}$ is defined as
\begin{equation}
G_{0}=4 \pi \int_{\rm FUV} J_{\nu} d\nu \bigg/ 1.6 \times 10^{-3},
\label{eq:G0}
\end{equation}
where the integral is over the FUV radiation in the range
(5.12-13.6)~eV.  In the above formula (eq.~\ref{eq:peheat}), small
($<15$\AA) grains contribute about half of the heating.

Molecular hydrogen efficiently forms on dust grains by depositing the
chemical binding energy on the grain surface. Following Tielens and
Hollenbach (1985), we compute the H$_{2}$ formation coefficient as,
\begin{equation}
k_{\rm gr}=\frac{6.0 \times 10^{-17}(T/300 {\rm K})^{1/2} f_{\rm a} 
Z/Z_{\sun}} 
{ 1+4.0 \times 10^{-2}(T+T_{\rm gr})^{1/2}
   +2.0 \times 10^{-3} T +8.0 \times 10^{-6} T^{2}}
\end{equation}
where
\begin{equation}
f_{\rm a}=
[1+ {\rm exp}(7.5 \times 10^{2}(1/75 -1/T_{\rm gr}))]^{-1}.
\end{equation}

\section{Results}
\label{sec:results}

In what follows, we will first discuss the results obtained for
the thermal evolution of metal--free gas clouds, 
and then describe the effects induced by the presence of metals and dust grains.

\subsection{Metal--free Clouds}
\label{sec:metal-free}

The thermal evolution of metal--free clouds 
irradiated by a FUV radiation background is expected to change with radiation temperature 
$T_{\ast}$ and intensity $J_{21}$. 
The models with a radiation temperature of $T_{\ast}=10^{4}$~K 
($10^{5}$~K) are shown in Figure \ref{fig:nT.T14} 
(\ref{fig:nT.T15}, respectively) for different values of intensity $J_{21}$.

Initially, i.e. at low densities, the temperature increases
adiabatically, because there is not enough H$_2$ to activate cooling.
In the no radiation case, when the density is $\sim 1{\rm cm^{-3}}$
and the temperature is $\sim 1000$~K, sufficient H$_2$ is formed and,
as a result, the temperature decreases.  It is to be noted that the
relatively low temperature where this condition is met does not
contradict previous results (BL03).  In fact, the predicted
temperature of each fluid element in the simulation of BL03 shows a
large scatter at low densities.  This scatter reflects the radial
temperature gradient, and the central value, which we calculate here,
corresponds to the lower boundary of the scattered points and it is in
agreement with our result.  We expect that the central temperature of
the gas cloud does not reach the virial temperature of the host dark
matter halo since the innermost region starts to cool and collapse
during the adiabatic compression and does not experience the
virialization shock.

As the external radiation intensity $J_{21}$ increases, the onset of
H$_{2}$ cooling is delayed because higher densities and temperatures
are required for H$_2$ formation to compensate for the
photodissociation.  If the UV intensity is below a threshold value,
$J_{\rm 21, thr}$, which we find to be in the range $10^{2}-10^{3}$
for $T_{\ast}=10^{4}$~K and $(1-3)\times 10^{5}$ for
$T_{\ast}=10^{5}$~K, there is always a density at which H$_2$ cooling
starts to become effective.  The temperature then decreases and
eventually reaches the no--radiation evolutionary track, along which it
evolves thereafter.  On the other hand, if the radiation is stronger
than the threshold value, H$_2$ cooling never becomes important.  In
this case, atomic hydrogen cooling by H excitation (for $\la
10^{7}{\rm cm^{-3}}$) and H$^{-}$ free-bound (f-b) emission (for $\ga
10^{7}{\rm cm^{-3}}$), are the main cooling channels (see Figure
\ref{fig:cool.T14J13D00M00}).

In Figure \ref{fig:nT.T14}, runs with higher initial temperature (210~K) are 
also shown (dotted lines). During the initial adiabatic phase, the temperature 
at a given density is proportional to its initial value, and thus higher in runs 
with higher initial temperature.
However, after the onset of efficient radiative cooling, these initially
different thermal evolutionary tracks soon converge.
At higher densities, the results are independent of the initial temperature (see Figure \ref{fig:nT.T14}).

As it can be inferred from Figs. \ref{fig:nT.T14} and \ref{fig:nT.T15},  we find that the threshold value, 
$J_{\rm 21, thr}$, is lower for a radiation temperature of $T_{\ast}=10^{4}$~K than for
$T_{\ast}=10^{5}$~K.  Thus, for comparable radiation intensities,
$J_{21}$, the lower $T_{\ast}$ radiation has a stronger impact on the
cloud evolution.  To understand why this is the case, 
in Figure \ref{fig:radparam} we show
the H$_2$ and H$^{-}$ photodissociation rates, for the same intensity $J_{21}=1$.
The dilution factor $W$, defined by $J_{\nu}\equiv W
B_{\nu}(T_{\ast})$, which was used in Omukai \& Yoshii (2003), is also
shown for reference.  
As the figure shows, the H$^{-}$ photodissociation rate decreases
steeply with $T_{\ast}$, while the H$_2$ photodissociation rate
remains nearly constant.  The H$_2$ and H$^{-}$ photodissociation rate
coefficients are
\begin{equation}
k_{\rm H_2 ph}=1.4 \times 10^{9} J_{\rm \nu}(12.4 {\rm eV}) 
\end{equation}
in the unattenuated case and
\begin{equation}
k_{\rm H^{-} ph}= \int_{0.755{\rm eV}} \frac{4 \pi J_{\nu}}{h \nu} 
\sigma_{\nu} d\nu
\end{equation}
where the lower limit on the latter integral, 0.755 eV, is the
threshold energy for H$^{-}$ photodissociation. Since the radiation
field is normalized at the Lyman limit (13.6~eV), $k_{\rm H_2 ph}$ is
not sensitive to $T_{\ast}$, whereas $k_{\rm H^{-} ph}$, which
reflects the radiation field above 0.755~eV, depends significantly on
the adopted $T_{\ast}$.  H$_2$ formation proceeds via a two step
process (H$^{-}$ channel),
\begin{equation}
{\rm H} + e \rightarrow {\rm H^{-}} + \gamma, 
\label{eq:reaction1}
\end{equation}
and  
\begin{equation}
{\rm H^{-}} + {\rm H} \rightarrow {\rm H_2} + e . 
\label{eq:reaction2}
\end{equation}
If H$^{-}$ is photodissociated, it can not activate the
second step (\ref{eq:reaction2}). This is the reason why
a lower $T_{\ast}$ radiation field leads to a less efficient
H$_2$ formation rate and to a lower H$_2$ fractional abundance 
than a higher $T_{\ast}$ radiation field.

Cooling via H$^{-}$ f-b emission occurs through the radiative
association reaction (\ref{eq:reaction1}). Subsequently, H$^{-}$ is
collisionally dissociated through the reaction
\begin{equation}
\label{eq:reaction3}
{\rm H^{-}} + {\rm H} \rightarrow 2 {\rm H} + e,  
\end{equation}
and the whole process results in a net cooling by the emitted photon.
Thus, as long as the collisional dissociation rate exceeds the
photodissociation rate, the H$^{-}$ f-b emission is hardly quenched,
even in the presence of a strong FUV radiation field.
Due to the small opacity, H$^{-}$ f-b cooling becomes important
only at high densities ($\ga 10^6 {\rm cm^{-3}}$) 
and temperatures ($\sim$ several $10^{3}$~K), where the above 
condition is always satisfied.
Then, H$^{-}$ f-b cooling is not affected by photodissociation.

As shown in Figure \ref{fig:cool.T14J13D00M00}, radiative cooling and 
compressional heating rates almost balances for a wide range of densities.
Namely, in equation (\ref{eq:energy}), the two terms in the right hand side
are almost cancels and the left hand side is much smaller than those terms.
Suppose that the radiative cooling rate per unit volume 
depends on the density and temperature as
\begin{equation}
\Lambda_{\rm rad}/\rho_{\rm gas} \propto \rho_{\rm gas}^{\alpha-1}  
T^{\beta}.
\end{equation}
Since the compressional heating rate 
\begin{equation}
-p\frac{d}{dt}\frac{1}{\rho_{\rm gas}}= \frac{p}{\rho_{\rm gas} t_{\rm col}} 
\propto \rho_{\rm gas}^{1/2} T,
\end{equation} 
the thermal balance of those two terms results in the following 
temperature evolution: 
\begin{equation}
T \propto {\rho_{\rm gas}}^{\frac{3/2-\alpha}{\beta-1}}.
\label{eq:Trho}
\end{equation}
Both the H line and H$^{-}$ f-b emissions are very sensitive to 
temperature, and thus $\beta >1$.
For those collisional processes, $\alpha =2$ for fixed
chemical abundances. 
With chemical evolution, it deviates from 2, but remains $>3/2$.
Therefore, the exponent in equation (\ref{eq:Trho}) is negative for 
the atomic-cooling track as long as the cloud is optically thin:
the temperature decreases with density as observed in Figures 
\ref{fig:nT.T14} and \ref{fig:nT.T15}.
On the other hand, on the molecular-cooling track, $\alpha =1$ for
densities higher than the critical value for the LTE.
Thus, the temperature increases with density for 
$n_{\rm H}\ga 10^{4} {\rm cm^{-3}}$.  

The existence of a threshold UV background and the discontinuity of 
thermal evolution at this value are due to the presence of 
non-local thermodynamic equilibrium (non-LTE) to LTE transition 
of H$_2$ ro--vibrational level population at 
$\sim 10^{4} {\rm cm^{-3}}$.
When the gas density is higher than this value, the cooling rate 
saturates and more H$_2$ is needed to compensate for compressional heating.
In addition, after the LTE is reached, 
collisional dissociation rate is enhanced owing to 
a large H$_2$ level population in the excited levels.
Thus, if a strong FUV radiation delays H$_2$ formation and cooling 
until the critical density for LTE is reached, 
a fraction of the remaining H$_2$ is collisionally dissociated.
Thus the gas cloud is no longer able to cool by H$_2$ even at a
later phase of the evolution.
On the other hand, if the UV background is slightly smaller 
than the threshold, the cloud begins to cool by H$_2$ and the 
temperature begins to fall before the collisional dissociation 
effect becomes significant (see Figure 5b in Omukai 2001 
for cooling rates by each process in such a case).
The lower temperature allows further H$_2$ formation 
and resultant cooling. 
The cooling proceeds in this accerelated fashion 
and the temperature eventually reaches the molecular cooling track.
This is the origin of the dichotomy between the atomic and molecular
cooling tracks.
To summarize, the main effect of the FUV radiation is to 
photodissociate H$_2$ directly and to decrease the
H$_2$ formation rate through photodissociating H$^{-}$.
If these two processes inhibit H$_2$
formation and cooling until the critical density for LTE is reached,
the gas remains warm ($\ga$ several thousands K) and H$_2$ is 
collisionally dissociated at higher densities.
Thus the high density evolution is not affected by the presence 
of the FUV field and depends only on the temperature at the H$_2$ 
critical density. 

\subsection{Metallicity Effects on Irradiated Clouds}
\label{sec:metal}

In this section, we show the effects induced by the presence of metals
and dust grains on the thermal evolution of gas clouds irradiated by a
FUV field with a mean intensity larger than $J_{21,\rm{thr}}$.  In
what follows, the total metallicity is expressed relative to the solar
value, as ${\rm [M/H]} \equiv {\rm log}(Z/Z_{\sun})$.  Unless
specified otherwise, the fractions of metals in the gas phase and in
dust grains are assumed to be the same as in the interstellar medium
(ISM) of the Galaxy. Specifically, the number fractions of C and O
nuclei in the gas phase with respect to H nuclei are 
$y_{\rm C,gas}=0.927 \times 10^{-4}Z/Z_{\odot}$ and 
$y_{\rm O, gas}=3.568 \times 10^{-4}Z/Z_{\odot}$.
The mass fraction of dust grains relative to the mass in gas 
is $0.939 \times 10^{-2}Z/Z_{\odot}$ below the
ice-vaporization temperature ($T_{\rm gr} \la 100$~K).

In Figure \ref{fig:nT.JZ} we present the thermal evolution of clouds
with metallicity in the range $-6 \le $[M/H]$\le -3$ irradiated by
extremely strong FUV radiation fields. The parameters of the radiation
fields are
(a) $T_{\ast}=10^{4}$~K, $J_{21}=10^{3}$ and 
(b) $T_{\ast}=10^{5}$~K, $J_{21}=3 \times 10^{5}$, respectively.
Under these conditions, the clouds would collapse only via atomic
cooling in the absence of metals or dust grains (see
Figs.~\ref{fig:nT.T14} and \ref{fig:nT.T15}).  For a metallicity as
low as [M/H] $\la -6$, the predicted thermal evolution follows the
metal--free track. In both panels of Figure \ref{fig:nT.JZ}, deviations
from the metal--free tracks start to appear at a density $\sim
10^{11}{\rm cm^{-3}}$ when the metallicity is [M/H]$\simeq -5.3$.  
For the sake of comparison, thin lines show the expected evolution
in the absence of radiation for the same initial values of metallicity. 
At metallicity [M/H]$= -5.3$, although the temperature drops and 
eventually reaches the molecular-cooling track 
at $\sim 10^{16}{\rm cm^{-3}}$, this arrival is 
after the minimum in the molecular cooling at $\sim 10^{14}{\rm cm^{-3}}$.
With a slightly higer metallicity of [M/H]$= -5$, this arrival 
takes place at $\sim 10^{11-12}{\rm cm^{-3}}$, 
and the temperature subsequently decreases
to the minimum in the no-radiation case.
For higher metallicities, the temperature drop occurs at lower density 
and the temperature minima becomes lower.
In Figure \ref{fig:cool.2} we show the cooling
and heating rates contributed by each process during the evolution of
the cloud with $T_{\ast}=10^{4}$~K, $J_{21}=10^{3}$ and [M/H]=-5. 
Up to $10^{10}{\rm cm^{-3}}$, cooling is dominated by the 
H line emission (denoted as ``H'' in the Figure; 
$\la 10^{7}{\rm cm^{-3}}$) and 
H$^{-}$ f-b emission (``H$^{-}$ f-b''; 
$\ga 10^{7}{\rm cm^{-3}}$), and 
the cloud collapses along the atomic cooling track 
(see Fig.\ref{fig:nT.JZ} a).
However, at a density $\sim 10^{10}{\rm cm^{-3}}$, 
cooling by the dust grain (``grain'') 
becomes dominant and causes 
the sudden temperature drop.
Now the temperature is lower than that in the atomic cooling track,
the H$_{2}$ collisional dissociation rate is also reduced, which
causes a high equilibrium value of the H$_2$ fraction.  As a result of
H$_2$ cooling, the temperature decreases further, although this effect
is almost completely balanced by heating due to H$_2$ formation
(``H$_2$ form'').  Note that fine-structure line cooling (``CII, OI'')
is not important at such low metallicities (see the discussion below).
Eventually, the thermal evolutionary tracks reach
those of the corresponding metallicity in the no--radiation case (shown
as thin curves in Fig.\ref{fig:nT.JZ}) and evolve along them
thereafter.

We also run models with an external UV field with parameters
$T_{\ast}=10^{4}$~K, $J_{21}=10^{4}$, 
which is 10 times stronger than that considered in Fig.\ref{fig:nT.JZ} (a) 
and we found that the critical value of the metallicity at which deviations
from the metal--free evolution appear, $Z_{\rm cr} \sim 5 \times
10^{-6} Z_{\odot}$ does not depend on the intensity of the FUV
radiation as long as $J_{21} > J_{\rm 21, thr}$.
Furthermore, for the same value of the metallicity, the
evolutionary tracks at high densities ($\ga 10^{4}{\rm cm^{-3}}$) are
independent of the type of external radiation, as can be seen
comparing the results in panels (a) and (b) of Figure \ref{fig:nT.JZ}.  
In fact, once the evolution has reached the density at which 
molecular and atomic cooling tracks bifurcate ($\ga 10^{4}{\rm
cm^{-3}}$), collisional processes rather than radiative ones dominate
the energy balance (see also \S~\ref{sec:metal-free}).

It is interesting to stress that the physical processes responsible
for the origin of a critical metallicity and its numerical value are
the same as those found in the absence of FUV radiation (Schneider et
al. 2003; Omukai et al. 2005). This is because despite the higher gas
temperatures induced by the presence of a strong FUV field (several
thousands of degrees), the dust temperature remains at a few tens of
degrees until the energy transfer rate from gas to dust by collisions
become important and the dust and gas temperatures approach each other
(with the associated gas cooling).
Nevertheless, the ultimate fate of protogalaxies at low metallicity 
can be significantly affected by the presence of the UV
flux (see discussion in \S~\ref{sec:corecollapse} below).  
In Figure~\ref{fig:nT.Tgr}, the evolution of dust and gas temperatures is
shown for the lowest metallicity tracks presented in
Figure~\ref{fig:nT.JZ} (a). 
The disappearance of the dust temperature curve for
${\rm [M/H]}=-6$ at a density of $\sim 10^{14}{\rm cm^{-3}}$ is
due to complete vaporization of grains, which occurs at $\simeq
1300$~K.  At dust temperatures higher than this value, grains are no
longer present in the cloud.
Other smaller discontinuities in the dust temperature also reflect
vaporization of some dust compounds.  For example, those at $\simeq
130$~K (at $\sim 10^{9}{\rm cm^{-3}}$ for ${\rm [M/H]}=-6$ and -5; at
$\sim 10^{11}{\rm cm^{-3}}$ for ${\rm [M/H]}=-4$) are due to
vaporization of water ice.
This figure indeed shows that despite the high gas temperature, 
the dust temperature remains low at a few 10K, which allows 
survival of grains until very high densities.

As discussed above, OI and CII line
emission contributes negligibly to gas cooling in the metallicity and
density range where the effects of dust grains start to become
relevant ($Z_{\rm cr} \sim 5 \times 10^{-6} Z_{\odot}$, $n_{\rm H}
\sim 10^{10}{\rm cm^{-3}}$).
Metal--line cooling causes a deviation from the metal--free atomic track only 
when the metallicity reaches ${\rm [M/H]} \sim -3$.
In this case, since the temperature track converges to the $J=0$ track 
before the dust--cooling phase, two temperature minima appear at 
$10^{5}{\rm cm^{-3}}$ and $10^{10}{\rm cm^{-3}}$ 
(see Fig.\ref{fig:nT.JZ} a, b).
In the absence of dust grains, a
higher fraction of metals is required to cool the gas at a rate such
that the thermal evolution deviates from the atomic cooling metal--free
tracks. 
To demonstrate this, we have performed a numerical
experiment where we have suppressed the contribution of dust grains to
the energy balance of the collapsing clouds. 
The results for models with radiation field parameters
of $(T_{\ast}=10^{4}$~K, $J_{21}=10^{3})$ are shown in 
Figure~\ref{fig:nT.gasZ}.
When the metallicity is below ${\rm [M/H]} \simeq -3.5$, 
the temperature evolution is exactly the same as the metal-free one  
(shown by the  $2 \times 10^{-4}Z_{\sun}$ track in the Figure).
For higher metallicity, fine-structure line cooling 
becomes dominant when $n_{\rm H} \la 10^{4}{\rm cm^{-3}}$
and the temperature drops abruptly by more than two orders of magnitude.
Therefore we find that the critical metallicity 
${\rm [M/H]} \simeq -3.5$ ($\simeq 3 \times 10^{-4} Z_{\sun}$) 
required to modify the thermal evolution is almost two orders of magnitude 
higher than in models with dust.  This
level of metallicity is approximately the same as the value at which
metal cooling rate exceeds the H$_2$ cooling rate in clouds which
are already collapsing by molecular cooling (Bromm \& Loeb 2003b,
Santoro \& Shull 2006; Frebel, Johnson \& Bromm 2007).  
 
\subsection{Fragmentation Properties}
\label{sec:fragmentation}

The thermal properties of star--forming clouds have an important
influence on how they fragment into stars (Larson 2005). There is
observational evidence that proto--stellar cores have a mass spectrum
which resemble the stellar initial mass function (IMF), indicating
that cloud fragmentation must be responsible for setting some
fundamental properties of the star formation process 
(e.g., Motte, Andre, \& Neri 1998; Lada et al. 2007; 
for the recent reviews, see Bonnell, Larson \& Zinnecker
2006 and Elmegreen 2008).
 
Roughly speaking, fragmentation occurs efficiently when the 
effective adiabatic index
$\gamma \equiv \partial {\rm ln} p/\partial {\rm ln} \rho \la 1$, 
i.e., during the temperature drops,
and almost stops when isothermality breaks ($\gamma \ga 1$) 
as also shown by the simulations
of Li, Klessen \& Mac Low (2003). Thus, consistent with Schneider et
al.  (2002, 2003, 2006) we can adopt the density at which the equation
of state first becomes softer than $\gamma \approx 1$ to identify the
preferred mass scale of the initial fragments (Inutsuka \& Miyama
1997, Jappsen et al. 2005). The fragment mass is given roughly by the
Jeans mass (or Bonnor--Ebert mass) at this epoch,

\begin{equation}
M_{\rm frag} = M_{\rm J}(n_{\rm frag}, T_{\rm frag}) \propto T_{\rm frag}^{3/2} \, \, n_{\rm frag}^{-1/2}.
\label{eq:mfrag}
\end{equation} 

In the absence of an external FUV radiation field, the temperature of
metal--free clouds decreases with density in the range $1 {\rm cm^{-3}}
\la n_{\rm H} \la 10^4 {\rm cm^{-3}}$ and increases at higher
densities, after the major coolant H$_2$ has reached the LTE. Dense
cores form around this density with typical masses of
$10^{3}M_{\sun}$, which is close to the Bonnor--Ebert mass at this
thermal state (Bromm, Coppi, \& Larson 1999, 2002; Abel, Bryan, \&
Norman 2002). As the metallicity increases to $Z_{\rm cr} = 10^{-6}
Z_{\odot}$, dust--induced fragmentation leads to solar or sub--solar
fragments (Schneider et al. 2006), making a fundamental transition in
the characteristic mass scales of proto--stellar cores.

It is important to stress that the presence of a temperature dip in
the thermal evolution, and the softening of the equation of state
$\gamma <1$, imply only the possibility of fragmentation. For example,
fragmentation depends also on the initial conditions, and requires the
existence of sufficiently large initial density perturbations.  In the
first cosmological objects, which are barely able to cool and
collapse, fragmentation can be less efficient. Still, self--gravitating
cores of mass comparable with that predicted by the above criterion
are observed to form in high--resolution 3D simulations (Abel, Bryan,
\& Norman 2002, Yoshida et al. 2006).  Even for turbulent molecular
clouds of solar metallicity, 3D simulations show that fragmentation is
efficient when $\gamma\approx 0.7$ and it is suppressed after $\gamma$
increases to $\approx 1.1$ (Jappsen et al. 2005).

The evolution with density of the effective adiabatic index, $\gamma$, 
is presented in Figure \ref{fig:gamma} for clouds with initial
metallicities [M/H]=$-\infty$, -6, -5.3, -5, -4, and -3,  
irradiated by a field 
with parameters $T_{\ast}=10^{4}$K and $J_{\ast}=10^{3}$, 
whose temperature evolution is shown in Fig. \ref{fig:nT.JZ} a.
The application of the above arguments to predict the typical fragment
mass from the thermal evolution of the clouds is not straightforward 
because, along the metal--free atomic cooling tracks and over a broad 
density range $10^{1-16}{\rm cm^{-3}}$, the effective adiabatic index remains
$\gamma \simeq 1$, although slightly below unity (0.95 - 1; see Fig. 
\ref{fig:gamma}, top panel).
If we adopt $\gamma_{\rm frag}=1$ as the threshold value of the 
effective adiabatic index for fragmentation, in this case
fragmentation would be expected to occur up to densities
of $\sim 10^{16}{\rm cm^{-3}}$, leading to solar--mass fragments, as
discussed by O2001 and Omukai \& Yoshii (2003).  In contrast, the
numerical simulations by BL03 show that down to the highest density
reached by the simulations ($\la 10^{9} {\rm cm}^{-3}$) fragmentation
is very inefficient. Even with some degree of rotation, the cloud
fragments at most into two pieces, resulting in a binary
system. Although fragmentation might occur at higher densities, in
BL03's calculations neither efficient fragmentation leading to the
formation of a star cluster, nor the growth of elongation of the
clouds is observed.  
We speculate that this result is due to the following reasons.  
The objects considered by
BL03 are those only marginally able to collapse by atomic cooling, and
thus are initially close to the hydrostatic equilibrium. During this
initial epoch, the Jeans mass is large, and density and velocity
perturbations are erased by pressure forces.
In addition to this little initial seed perturbation, 
since $\gamma$ is only slightly below unity, 
the growth of perturbation would be very slow.
Thus the perturbation might not grow enough to cause fragmentation. 

Note however that, although we find that along the atomic cooling
tracks H$^{-}$ cooling is the dominant cooling agent at high
densities, $\ga 10^{7}{\rm cm^{-3}}$, this process is not considered
in the simulation of BL03, which implements only H Ly$\alpha$ cooling.
To check whether this omission might cause the lack of fragmentation,
we have followed the evolution of a metal--free cloud under the
influence of an external FUV radiation field with $T_{\ast}=10^{4}$K
and $J_{21}=10^{3}$ but turning off the H$^{-}$ cooling by hand.  The
result is shown in Figure \ref{fig:Hmbf}. With no H$^{-}$ cooling, the
cloud follows a slightly higher temperature track when the density is
$\ga 10^{7}{\rm cm^{-3}}$.  However, below $\sim 10^{12} {\rm
cm^{-3}}$, the difference is small and it has a weak effect on the
cloud dynamics. Therefore the inclusion of H$^{-}$ cooling would not
affect the results of BL03's simulation, which is limited to densities
$<10^{9}{\rm cm^{-3}}$.

On the basis of these considerations, we assume that for metal--free
clouds irradiated by a strong FUV background, fragmentation does not
occur during the atomic--cooling phase, where $\gamma \simeq 1$, and it
occurs only when the temperature drops more rapidly, where 
$\gamma < \gamma_{\rm frag} <1$, by molecular
cooling, that is when $J_{\nu} < J_{\nu,{\rm thr}}$.

In the metal--enriched, irradiated clouds we studied, the temperature
dip due to dust cooling occurs at very high densities, $10^{10} {\rm
cm^{-3}}$, deep in the interior of the collapsing clouds, where
pre--existing density perturbations might also be erased by pressure
forces. However, in this regime we find $\gamma \la 0.5$ as shown 
in Figure \ref{fig:gamma};
fragmentation has also been confirmed to occur in two independent
hydrodynamic simulations of collapsing clouds not irradiated by
external FUV fields (Tsuribe \& Omukai 2006, Clark, Glover, \& Klessen
2008). Therefore, we expect that for metallicities ${\rm [M/H]} \ga
-5$, when the thermal evolutionary tracks shown in
Figure~\ref{fig:nT.JZ} suddenly deviate from the atomic--cooling track,
in other words, when $\gamma$ falls sufficiently below unity 
(Figure \ref{fig:gamma}), 
the clouds begin a vigorous fragmentation, which then lasts until the
temperature increases again. 
The value of $\gamma_{\rm frag}$ to cause fragmentation is uncertain 
as discussed above, but likely to be slightly below unity.
In the following, for the sake of definiteness, adopt $\gamma_{\rm frag}=0.8$ as the 
fiducial value below which fragmentation is triggered 
(the lower horizontal lines in Fig. \ref{fig:gamma}), 
and use it to define the properties of the fragments. 
This choice refrects the fact that for $\gamma=0.7$ 
efficient fragmentation has been observed to occur in the numerical simulations
of Jappsen et al. (2005).
In all cases, once molecular cooling becomes efficient, 
$\gamma$ soon falls below $\la 0.5$ (Fig. \ref{fig:gamma}). 
Varying the threshold value $\gamma_{\rm frag}$, say, 
by $\sim 0.1$, leads to fragmentation densities whose 
differences are within an order of magnitude.
For example, when [M/H]$=-4$ fragmentation begins 
at ${\rm log}~n_{\rm H} ({\rm cm^{-3}})= 7.5, 8$, and 8.3
for $\gamma_{\rm frag} = 0.7, 0.8$, and $0.9$, respectively.
We assume that fragmentation stops when $\gamma$ exceeds unity again.
For [M/H]$=-4$, this occurs at $n_{\rm H} \sim 10^{13}{\rm cm^{-3}}$.

The mass scale of the final fragments is
given by the Jeans mass at the temperature minimum, i.e., 
when $\gamma$ exceeds unity.  When the initial
metallicity is ${\rm [M/H]} \simeq -5$, the temperature minimum
corresponds to $300$~K at $n_{\rm H} = 10^{14} {\rm cm}^{-3}$, and
thus the typical fragment mass is $0.1 M_{\sun}$.  As the metallicity
increases, both the density and temperature at the fragmentation scale
decrease, being ($10^{13} {\rm cm}^{-3}$, 150K) for ${\rm [M/H]}
\simeq -4$, and ($10^{11} {\rm cm}^{-3}$, 30K) for ${\rm [M/H]} \simeq
-3$. However, the corresponding fragment mass scale remains $\sim
0.1 M_{\sun}$, because the variations of density and temperature almost
cancel out (see eq.~\ref{eq:mfrag}).
In some cases, e.g. ${\rm [M/H]} \sim -3$ gas in a $J_{21}>J_{\rm 21, thr}$ 
field, two fragmentation epochs (${\rm log}~n_{\rm H}=3.5-5.1$ and $8.5-10.7$ 
for ${\rm [M/H]}=-3$) appear, which 
corresponds to two dips in the temperature (or $\gamma$) evolutionary track.
The outcome of this kind of track is not clear without any numerical 
work studying their effect.
Here, we speculate that the first dip produces clumps
as a result of the fragmentation of clouds.
Then the clumps fragments again into cores owing to the second dip. 

In the absence of dust, the temperature minimum appears at a lower
density, $n_{\rm H} \sim 10^{5}{\rm cm^{-3}}$, and higher metallicity
${\rm [M/H]} \simeq -3.5$ (see Fig.~\ref{fig:nT.gasZ}).  Therefore,
the corresponding fragment mass remains as high as $10-100 M_{\sun}$
and the formation of sub--solar mass fragments is not possible in this
case. This property of pre--stellar clouds enriched only by gas--phase
metals has been already proven to hold in the absence of external FUV
fields (Schneider et al. 2006).

To summarize, our results show that in the presence of a sufficiently strong FUV radiation field 
the collapse of metal--free clouds by molecular cooling is inhibited and it can proceed only via 
atomic cooling. Under these conditions, cloud fragmentation is highly inefficient, leading at most 
to the formation of a binary system. The typical mass of pre--stellar clouds is therefore 
$~10^{5-6} M_{\odot}$
and the formation of a super massive star, seed of a super massive
black hole, is the likely outcome of the evolution (BL03). However,
this scenario is altered as soon as trace amounts of metals and dust
grains are present in the collapsing clouds: dust cooling leads to
fragmentation of the clouds into sub--clumps with mass as low as $\sim
0.1 M_{\odot}$ already at a floor metallicity of $Z_{\rm cr} \sim 5
\times 10^{-6} Z_{\odot}$. This conclusion holds independently of the
intensity and spectrum of the FUV radiation field. In the absence of
dust, an enrichment level of $Z_{\rm cr} \sim 3 \times 10^{-4}
Z_{\odot}$ is required for OI and CII line cooling to fragment the
cloud; the fragments in this case are predicted to be relatively more
massive, $\sim 10 - 100 M_{\odot}$.

\subsection{Dynamical Interactions and Accretion}
\label{sec:corecollapse}

Since dust--induced fragmentation takes place at high densities, a
dense proto--stellar cluster is expected to form (Omukai et al. 2005,
Schneider et al. 2006, Clark et al. 2008).  As an example, when the
initial metallicity of the collapsing cloud is ${\rm [M/H]}=-5$, the
sudden temperature drop, where $\gamma < \gamma_{\rm frag}=0.8$,  
begins at $T_{\rm drop} \sim 3500$~K and
$n_{\rm drop}\simeq 10^{10.5}{\rm cm^{-3}}$. At this stage, the size and
mass of the cooling region, or proto--cluster, are given by the
corresponding Jeans length, $\lambda_{\rm J} \simeq 4\times 10^{-3}$~pc, 
and mass $M_{\rm cl} \simeq 70 M_{\sun}$.
When a different threshold value $\gamma_{\rm frag}$ is adopted,
these quantities change, e.g., to $\lambda_{\rm J} \simeq 3\times 10^{-3}$~pc, 
and $M_{\rm cl} \simeq 40 M_{\sun}$ for $\gamma_{\rm frag}=0.7$ (0.9, 
respectively).
In the following order of magnitude estimation, we use 
$\lambda_{\rm J} \sim 1 \times 10^{-2}$~pc 
and $M_{\rm cl} \sim 100 M_{\sun}$ as typical values.
After virialization, the
proto--stellar cluster has a size half of this. Since each ultimate
fragment has a typical mass of $M_{\rm frag} \sim 0.1 M_{\sun}$, which
is set by the Jeans mass at the end of the fragmentation process
($n_{\rm H} \sim 10^{14}{\rm cm^{-3}}$), we expect that up to
$N_{\ast} \sim M_{\rm cl}/M_{\rm frag} \sim 1000$ low--mass star can be
formed and confined into a small region of size $\sim 0.01$~pc.  The
difference between the formation epochs of each protostar is of the
order of the free--fall time of the proto--cluster gas.  Since the
cluster begins to form in a dense cloud with density $\sim 10^{10}{\rm
cm^{-3}}$, protostar formation is synchronized on a timescale of
$\sim 300$ yrs.

The fate of dense, compact star clusters has been discussed
extensively in the literature (see, e.g. Rasio et al. 2004 for a
recent review, focusing on the possibility of intermediate BH, IMBH,
formation through a runaway collapse that is relevant in our case). It
is important to stress that, even assuming a star formation efficiency
of order unity (which seems likely when the density exceeds $\ga
10^{4}{\rm cm^{-3}}$; Alves, Lombardi \& Lada 2007), the stellar IMF
will be strongly affected by gravitational interactions, collisions
and mergers. In fact, observed properties of present--day star forming
regions, as well as numerical simulations, suggest that gravitational
fragmentation is probably responsible for setting a characteristic
stellar mass but the full mass--spectrum and the Salpeter--like slope
of the IMF are most likely formed through continued accretion and
dynamical interactions in a clustered environment (see Bonnell et
al. 2006 and references therein). Furthermore, in young and compact
star clusters supermassive stars may form through repeated collisions
(e.g. Portegies Zwart et al. 1999, 2004; Ebisuzaki et al. 2001).

We can therefore ask, what is the expected fate of the dense star
cluster forming in our clouds?  The evolution of a star cluster with
half--mass radius $R_{\rm cl}$ and mass $M_{\rm cl}$ proceeds on the
dynamical friction timescale (Binney \& Tremaine 1987),
\begin{equation}
t_{\rm fric} \simeq  
\frac{1.2\times 10^{5}}{\rm ln \Lambda}{\rm yr}
\left( \frac{r}{0.01{\rm pc}}\right)^{2}
\left( \frac{R_{\rm cl}}{0.01{\rm pc}}\right)^{-1/2}
\left( \frac{M_{\rm cl}}{10^{2}M_{\sun}}\right)^{1/2}
\left( \frac{m_{\ast}}{1M_{\sun}}\right)^{-1},
\end{equation}
which is the time required for a star with mass $m_{\ast}$, which is
on the massive side of the spectrum, to sink from the radius $r$ to
the cluster center by gravitational interactions with background, less
massive stars. 
In the following, the Coulomb logarithm ${\rm ln}\Lambda$ is taken to be $\simeq 7$, 
a value typical for open clusters.
In the above equation, the density profile is assumed
to be isothermal, $\rho \propto 1/r^{2}$.  
The dynamical friction timescale was originally derived 
for a fixed background.
What actually occurs for a cluster on this timescale 
is the equipartition of kinetic energy among the member stars.
Heavy stars move slowly and then drop deeper in the potential well,
leading to mass segregation in the cluster.
The stellar merger rate is
greatly enhanced if higher--mass stars reach the cluster center within
the lifetime of a very massive star, i.e. if $t_{\rm fric}$ is less
than a few Myr.

Using the estimated size and mass of the proto--cluster at the onset of
 dust--induced fragmentation for the [M/H] = -5 track, the dynamical
friction timescale for a star initially at radius $r$ can be rewritten 
more generically as
\begin{equation}
t_{\rm fric} \simeq
1.6\times 10^{3}{\rm yr}
\left( \frac{M_r}{10^{2}M_{\sun}}\right)^{2}
\left( \frac{\mu}{1.22} \right)^{3/2} 
\left( \frac{T_{\rm drop}}{5000{\rm K}}\right)^{-3/2}
\left( \frac{m_{\ast}}{1M_{\sun}}\right)^{-1},
\label{eq:tfric2}
\end{equation}
where $M_r$ is the mass enclosed inside radius $r$ and we have
expressed $M_{\rm cl}$ and $R_{\rm cl}$ in terms of $n_{\rm drop}$ and
$T_{\rm drop}$ (note that $n_{\rm drop}$ drops out of the equation).
Following Portegies Zwart et al. (2004), we assume that a very massive star can be
formed by stellar mergers if,
\begin{equation} 
t_{\rm fric} < 4 \, {\rm Myr}.
\label{eq:mergcond}
\end{equation}
From equation (\ref{eq:tfric2}), we infer that the inner region of mass 
\begin{equation}
M_{\rm fric}=5 \times 10^{3} M_{\sun}
\left( \frac{\mu}{1.22} \right)^{-3/4} 
\left( \frac{T_{\rm drop}}{5000{\rm K}}\right)^{3/4}
\left( \frac{m_{\ast}}{1M_{\sun}}\right)^{1/2}
\end{equation}
satisfies the condition (\ref{eq:mergcond}).  The mass fraction of
sinking stars relative to the cluster background stars, $f_{\rm
sink}$, is uncertain and probably depends on the stellar mass
spectrum, but can be safely assumed to be less than a half.  In
addition, not all the stars that sink to the center are incorporated
into the runaway merging object. Since merging events usually proceed
via three-- or more body interactions, where the runaway object and a
star coalesce by kicking the lightest star (Portegies Zwart et
al. 1999), the merging efficiency, $f_{\rm merg}$, with the runaway
object among stars fallen to the cluster center can be assumed to be
about a half.  Thus, the mass of the central object resulting from
this process can be estimated as,
\begin{eqnarray}
M_{\rm cen} &=& f_{\rm sink}f_{\rm merg}M_{\rm fric} \\
&\approx & 3.5 \times 10^{2} M_{\sun}
\left( \frac{f_{\rm sink}}{0.25}\right)
\left( \frac{f_{\rm merg}}{0.5}\right)
\left( \frac{T_{\rm drop}}{5000{\rm K}}\right)^{3/4}
\left( \frac{m_{\ast}}{1M_{\sun}}\right)^{1/2}. 
\label{eq:Mcen}
\end{eqnarray}

As can be seen in Figures \ref{fig:nT.JZ} and \ref{fig:gamma}, 
dust--induced fragmentation
in clouds with a metallicity ${\rm [M/H]}\la -4$, starts at $T_{\rm
drop} \sim 5000$~K, $n_{\rm drop} \ga 10^{8} {\rm cm^{-3}}$ and the
corresponding proto--cluster mass is $M_{\rm cl} <
10^{3}M_{\sun}$. That is, $M_{\rm cl} < M_{\rm fric}$ and the entire
cluster satisfies the condition (\ref{eq:mergcond}).  In this case,
the mass of the central object is limited by the mass of the cluster
rather than by $M_{\rm fric}$ and its final mass can be estimated as,
$M_{\rm cen} = f_{\rm sink}f_{\rm merg}M_{\rm cl} \la 100 M_{\sun}$.
On the other hand, dust--induced fragmentation of clouds with
metallicity, $-4 \la {\rm [M/H]} \la -3$, leads to proto--stellar
clusters with masses $M_{\rm cl} \ga M_{\rm fric}$ and the mass of the
central object is given by equation (\ref{eq:Mcen}). 
Thus, in this case a very massive star can form with $M_{\rm cen} 
\la 350 M_{\sun}$.
In either case, we note that the metal--poor, massive star ultimately
forming at the center of the halo is likely to leave behind a seed BH
remnant -- except in a narrow range of metallicity, where they produce
a pair instability supernova -- either by direct collapse or by 
fallback (Heger et al. 2003).
For higher metallicity, there are two episodes of fragmentation; 
a metal--induced one at low--density and a dust--induced one at high--density. 
In this case, the mass scale of the star cluster is relatively low 
and a massive BH seed is not formed. 

As pointed out above, the critical metallicity levels we find in the
case of strong FUV irradiation, at which the cloud behavior is
modified from the metal--free case, is very similar to the critical
metallicity found in earlier work for $J=0$.  
It is therefore important to ask whether the presence of the flux will, 
in fact, make any difference to the ultimate fate of the cloud.  
At metallicities above ${\rm [M/H]}\simeq -3$, the flux has essentially 
no impact on the evolutionary track of clouds at high densities. 
For example, comparing the thin and thick solid curves of 
Figure \ref{fig:nT.JZ} (a) it is clear that when 
${\rm [M/H]}= -3$, the flux has no effect at $n_{\rm H} \gsim 10^5{\rm cm^{-3}}$.  
At lower densities, we expect that the first fragmentation phase 
will occur as the Jeans mass drops from $M_J \sim 10^{6}~{\rm M_\odot}$ 
to $\sim 10^{3}~{\rm M_\odot}$ at $n_{\rm H} \sim 10^{3}{\rm cm^{-3}}$ in the 
$J=0$ case, and to $\sim 10^{2}~{\rm M_\odot}$ at $10^{5}{\rm cm^{-3}}$ 
in the strong UV background case. Therefore, the size of the molecular clumps 
that form is larger in the $J=0$ case. However, the thermal evolution 
thereafter is the same: the molecular clumps will experience a second 
phase of fragmentation when the Jeans mass falls further, from 
$M_J \sim 10~{\rm M_\odot}$ to $\sim 0.1~{\rm M_\odot}$.
Thus, for protostellar gas clouds with ${\rm [M/H]} \simeq -3$, 
the presence of an external UV field determines only the amount of gas 
in the envelope in which the $\sim 10~{\rm M_\odot}$ star cluster is embedded. 

In contrast, the presence of a FUV background significantly affects the
evolution of protostellar clouds with lower values of metallicity, 
${\rm [M/H]} < -3$. In these models, the atomic gas experiences only the
second phase of fragmentation induced by dust cooling, and therefore
more massive star clusters ($10^{2}-10^{3}M_{\sun}$, increasing with metallicity) 
are formed compared to those in the $J = 0$ limit (a few - $10M_{\sun}$, 
depending weakly on metallicity). In addition to the size of the star clusters, 
the mass and physical conditions of the corresponding envelopes are different:
when $J = 0$, the star cluster is embedded in a molecular envelope 
of $10^{3-4}~{\rm M_\odot}$ with temperature of a few 100~K, 
while, in the presence of an external UV field, the surrounding envelope 
is more massive ($10^{5-6}~{\rm M_\odot}$) and it is made by atomic gas 
at several 1000~K. As a consequence, the presence of a strong UV background 
case favors the formation of a more massive central star by stellar merger 
(larger stellar cluster mass and higher $T_{\rm drop}$ in eq. \ref{eq:Mcen}).

Finally, our results and the discussion above suggest that the direct
formation of a supermassive star or SMBH as massive as $\sim
10^{5-6}~{\rm M_\odot}$, as envisioned in the metal--free case, is not
possible when the metallicity is above a critical value, and the gas
fragments into smaller pieces.  However, we note that if fragmentation
of the inner regions of the collapsing protogalaxy is not fully
efficient, 
and if radiative and mechanical feedback from the stars does not expel 
the leftover gas from the nucleus, 
then the star cluster at the center of the halo, and its
coalesced massive remnant star, can be embedded within a thick
residual gaseous envelope with temperature $T_{\rm env} \sim $ several
thousand K.  Since this gas envelope is self--gravitating, with a
temperature below the virial temperature, it can undergo dynamical
collapse, and may still produce a SMBH either directly or by accretion
onto the central stellar--mass BH at the Bondi rate (see,
e.g. Begelman et al. 2006 and discussion therein).
In particular, in the UV irradiated case, the accretion rate onto 
the star cluster, and possibly to the central coalesced star, is 
high -- this is because of the high temperature in the envelope,
and since the accretion rate of the self--gravitating gas is given 
by (Shu 1977):
\begin{eqnarray}
\dot{M} &\simeq& \frac{c_{\rm s}^3}{G}\\
                  &=& 4\times 10^{-2} M_{\sun}/{\rm yr}
\left( \frac{T_{\rm env}}{5000{\rm K}} \right)^{3/2}.
\end{eqnarray}
In conclusion, higher initial mass, higher accretion rate, and 
larger amount of reservoir gas are more favorable for BH growth 
in the strong UV case than the case without radiation.

\section{Summary and Conclusions}
\label{sec:conclusions}

In this paper, we have investigated the thermal evolution
and fate of proto--stellar gas clouds in $T_{\rm vir} \ga 10^4$~K 
halos irradiated by a strong FUV background. Under these conditions,
which may apply to some dwarf galaxies collapsing close to the epoch of
reionization, we find that: 

\begin{itemize}

\item The effect of an external UV background is to photodissociate
H$_2$ directly and to decrease the H$_2$ formation rate through 
photodissociation of H$^{-}$. When the UV background reaches
a critical threshold value, $J_{\rm 21, thr}$, these two processes
inhibit H$_2$ formation and cooling until the critical density
for LTE is reached. Thereafter, the gas cloud can cool only
via atomic hydrogen transitions. 

\item For gas clouds of primordial composition, an
external UV background with intensity 
$J_{\rm 21} < J_{\rm 21, thr}$ only delays
the onset of H$_2$ formation and H$_2$ cooling 
becomes important at some (higher) density: fragmentation
occurs at densities $10^{3} {\rm cm^{-3}}-10^{5} {\rm cm^{-3}}$ 
leading to average fragment masses in the range $10^2 M_{\sun}-10^3 M_{\sun}$,
similarly to the case with $J_{\rm 21}=0$.

\item For $J_{21} > J_{\rm 21, thr}$, not enough H$_2$ is formed to
activate cooling and the evolution of primordial clouds is controlled
by atomic (H and H$^{-}$) cooling. The clouds collapse nearly
isothermally with a temperature of several thousands K (``atomic
track'') up to very high densities $\sim 10^{16} {\rm cm^{-3}}$.
According to previous numerical calculation by BL03, the clouds 
collapse directly into a single
$10^{5}-10^{6}M_{\sun}$ object, leading to super massive star and SMBH
formation. 
A core--envelope structure inevitably develops under
these circumstances, and the star grows by accretion onto an initially
small inner core. During the accretion phase, radiative and mechanical
feedback effects might become important and eventually halt the
accretion at some phase (e.g., Omukai \& Palla 2003; McKee \& Tan
2007).  If so, the mass of the central object can remain far below
$10^{5-6} M_{\sun}$.

\item Independently of the values and properties of the external FUV field (as long as
it is $J_{\rm 21} > J_{\rm 21, thr}$), deviations from the metal--free "atomic track"
start to appear when the gas is enriched by even trace amounts of metals and dust. 
When $Z> Z_{\rm cr} \simeq 5 \times 10^{-6} Z_{\odot}$, dust cooling induces fragmentation 
at $n_{\rm H} \sim 10^{10} {\rm cm}^{-3}$ and a proto--stellar cluster is expected to form with
average proto--stellar (fragment) mass of $\sim 0.1 M_{\odot}$. If only gas--phase metals are present,
a two orders of magnitude larger value of metallicity is needed, $Z_{\rm cr} \sim 3 \times 10^{-4} Z_{\odot}$,
before CII and OI line--cooling induce a deviation from the ''atomic track'', leading to 
fragmentation at $n_{\rm H} \sim 10^5 {\rm cm}^{-3}$ and to proto--stellar clusters with average
proto--stellar (fragment) mass of $10 - 100 M_{\odot}$. 

\item The physical processes responsible for the origin of a critical metallicity and 
of its numerical value are the same as those found in the absence of an external FUV field.
However, due to the higher gas temperature, the final outcome of the proto--stellar cloud
collapse can be significantly affected.
Namely, if we assume that the size of the proto--stellar cluster
formed by dust--induced fragmentation is set by the Jeans mass 
at the onset of the rapid temperature drop, 
it depends on the intensity
of the FUV background field: when $J_{\rm 21} < J_{\rm 21,thr}$, relatively
small clusters are formed (a few - 10$M_{\sun}$) whereas when 
$J_{\rm 21} > J_{\rm 21,thr}$, very dense star clusters with masses $100-1000~M_{\sun}$ 
are formed, at the center of which stellar coalescences are expected to occur.
The central merger object might grow to a very massive star of a few 
100 $M_{\sun}$.

\item In addition, the presence of an external FUV background affects the physical
conditions of the envelope surrounding the proto-stellar clusters: 
when $J_{\rm 21} < J_{\rm 21, thr}$ the envelope is fully molecular, with a mass
of $10^2 - 10^3 M_{\odot}$ and a temperature of a few $100$~K. Conversely, when
$J_{\rm 21} > J_{\rm 21,thr}$ the envelope is made of atomic gas and reaches a
mass of $10^5-10^6 M_{\odot}$ and temperature of several $1000$~K. The higher
temperature and larger gas reservoir favors BH growth by accretion, which can
be as high as $10^{-2} M_{\odot} {\rm yr}^{-1}$.  

\end{itemize}

According to the above, the conditions that would allow the formation
of the direct formation of a SMBH are (i) to be hosted within a
$T_{\rm vir} \sim 10^4$~K halo, which is ii) irradiated by a strong UV
field with $J_{21} > J_{\rm 21,thr}$, and (iii) still metal and dust
free, with $Z<Z_{\rm cr}$.  The main new result of the present paper
is that if the metalicity is too high, so that condition (iii) does
not hold, then instead of a SMBH, a dense cluster of low--mass star
forms at the halo nucleus. The stars in such a cluster may still
rapidly coalesce into a single massive star, which may produce an
intermediate--mass BH remnant, but with a smaller mass of $M\lsim
10^2-10^3~{\rm M_\odot}$.

While the above conclusion that even trace amounts of 
dust enable cooling and fragmentation of the proto--stellar
clouds appear to be robust, the exact value of the threshold 
metallicity is vulnerable to the uncertain nature of dust in 
early protogalaxies (see also Schneider et al. 2006). However,
it is interesting to note that when $Z \ge Z_{\rm cr}$ the 
characteristic fragment mass -- which is related to the
characteristic stellar mass -- is highly insensitive
to environmental conditions, such as the presence of an
external FUV radiation field, as also recently discussed by
Elmegreen, Klessen \& Wilson (2008).    

We warn the reader that our discussion on the 
nature and evolution of the resulting proto--stellar cluster 
is still speculative as it is based on a few numerical 
experiments which apply to dense stellar systems in 
present--day star forming regions (see the
discussion and references in section \ref{sec:corecollapse}). 
For example, we assume that the size of the cluster is set at 
the onset of the efficient cooling phase, which appears plausible
but has not yet been confirmed. In particular, in models with
$J_{\rm 21} > J_{\rm 21, thr}$ and $Z > 10^{-4} Z_{\odot}$ (when
both dust grains and gas-phase metals are present), 
the thermal evolution curves appear to have two separate 
temperature minima, which correspond to metal-- and dust-- induced
cooling and fragmentation. The fate of these collapsing clouds
is at present unknown, and dedicated numerical simulations 
would be highly desirable.
If proto-stellar clusters are indeed formed under the conditions that we
suggest, the formation and the nature of a central object by repeated
collisions and accretion is highly uncertain. For example, the fraction 
of stars on the massive side of the spectrum which falls to the cluster 
center (i.e., $f_{\rm sink}$ in eq. \ref{eq:Mcen}) by dynamical friction
is unknown and may vary significantly depending on the IMF.

Despite the above uncertainties, our results suggest that even trace
amount of metals preclude the rapid formation of SMBHs as massive as
$M\approx 10^{5-6}~{\rm M_\odot}$ in protogalactic halos.  While such
promptly appearing SMBHs would help solving the puzzle of the $M\gsim
10^{9}~{\rm M_\odot}$ quasar black holes at $z\gsim 6$, our results
suggest that the low--metallicity halos may instead produce dense
stellar clusters -- the cluster may coalesce to produce an IMBH, but
still with a much lower mass of $M\approx 10^{2-3}~{\rm M_\odot}$.

\acknowledgements This study is supported in part by the Grants-in-Aid
by the Ministry of Education, Science and Culture of Japan (16204012,
18740117, 18026008, 19047004:KO), by NASA through grant NNG04GI88G (to
ZH) and by the Pol\'anyi Program of the Hungarian National Office for
Research and Technology (NKTH).

\newpage

%%%%%%%%%%%%%%%%%%%%%%%%%%%%%%%%%%%%%%%%%%%%%%%%%%%%%%%%%%%%%%%%%% 
%%%%%%%    References   %%%%%%%%%%%%%%%%%%%%%%%%%%%%%%%%%%%%%%%%%%
%%%%%%%%%%%%%%%%%%%%%%%%%%%%%%%%%%%%%%%%%%%%%%%%%%%%%%%%%%%%%%%%%%

\newpage
\bigskip

%%%%%%%%%%%%%%%%%%%%%%%%%%%%%%%%%%%%%%%%%%%%%%%%%%%%%%%%%%%%%%%%%%
%%%%%%  (5) Table           %%%%%%%%%%%%%%%%%%%%%%%%%%%%%%%%%%%%%%
%%%%%%%%%%%%%%%%%%%%%%%%%%%%%%%%%%%%%%%%%%%%%%%%%%%%%%%%%%%%%%%%%%

\plotone{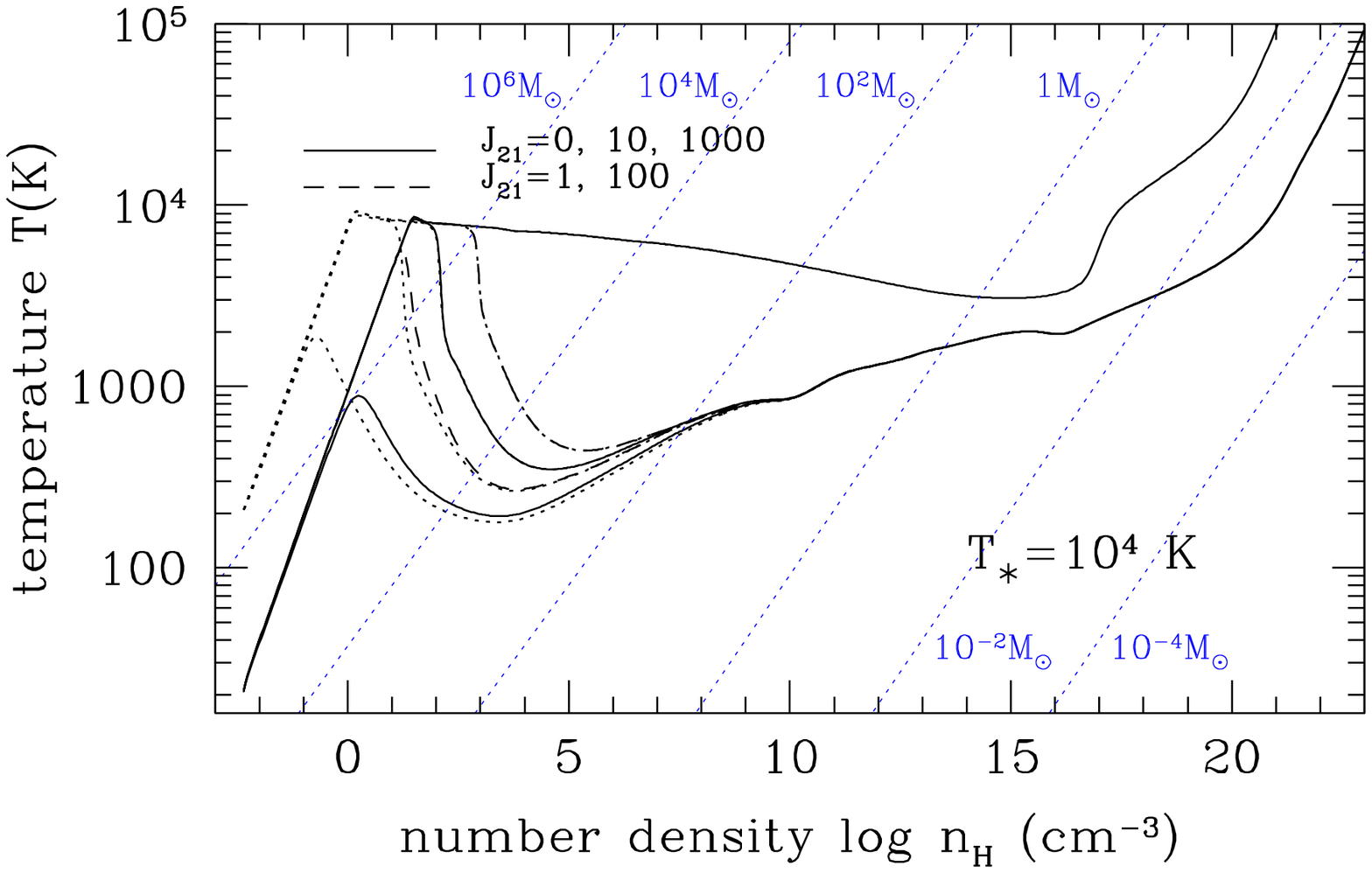} \figcaption[]{Temperature evolution of
metal-free clouds irradiated by a UV flux. The spectral shape is that
of a black--body spectrum with $T_{\ast}=10^{4}${\rm K}.  Models are
shown with FUV intensities at the Lyman limit of $J_{21}=0, 1, 10,
100$ and $10^{3}$, in the usual units of $10^{-21} {\rm
erg~cm^{-2}~sr^{-1}~s^{-1}~Hz^{-1}}$ (solid and dashed curvesfrom bottom to top; see the
legend in the panel). Diagonal dotted lines correspond to different
constant Jeans mass.
Models with higher initial temperature (210~K in contrast to 21 K 
in the fiducial models) are also shown by dotted lines. For those
with $J_{21}=10, 100$ and $10^{3}$, the temperature evolution 
at $n_{\rm H} \ga 10^{1.5}{\rm cm^{-3}}$ completely overlaps with that
predicted by the fiducial models.
\label{fig:nT.T14}}

\plotone{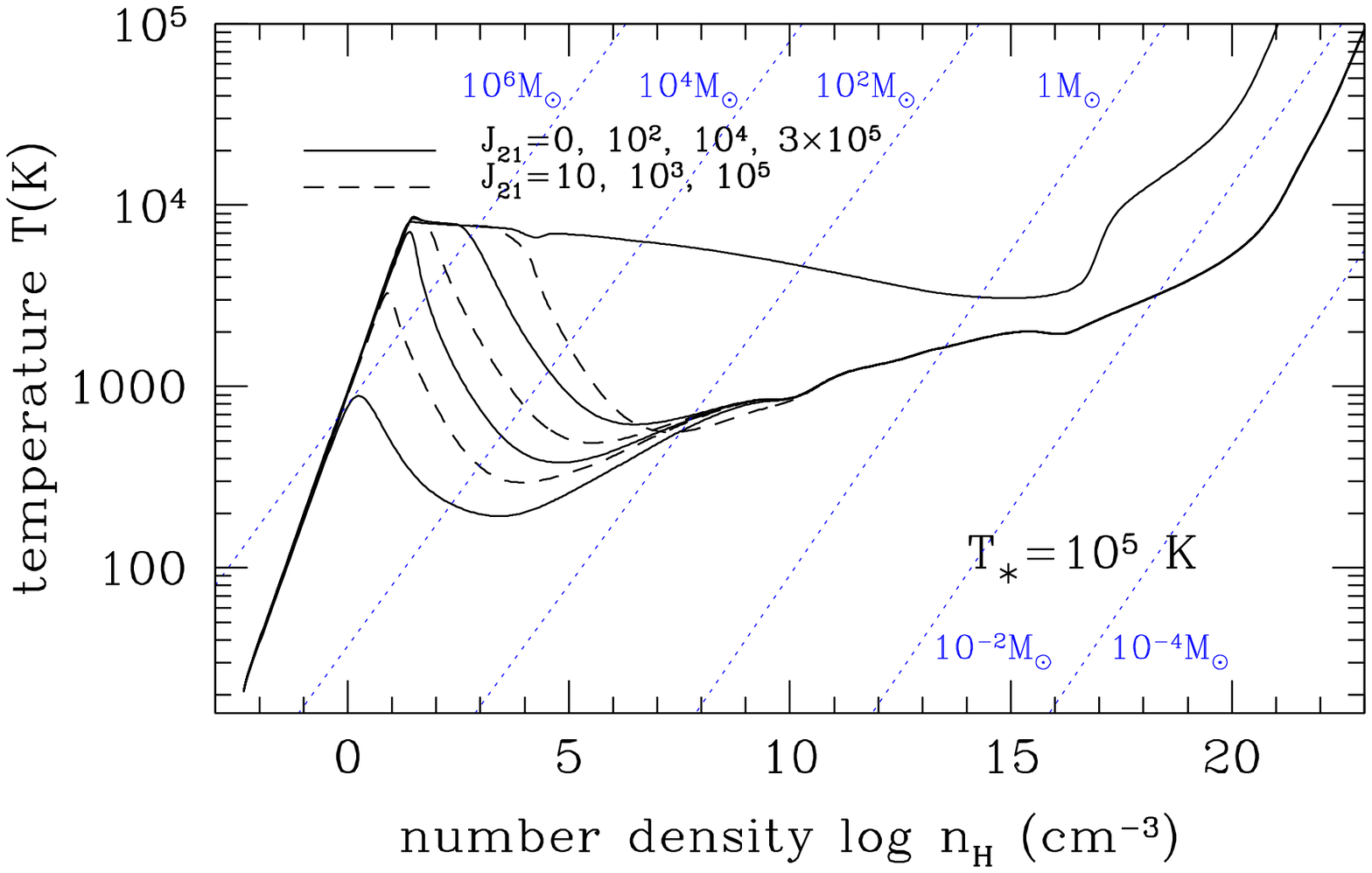} \figcaption[]{The same as Figure
\ref{fig:nT.T14}, except assuming a harder spectrum, with
$T_{\ast}=10^{5}${\rm K} and intensities $J_{21}=0, 10, 100, 10^{3},
10^{4}, 10^{5}$, and $3\times 10^{5}$ (solid and dashed curves from bottom to top).
\label{fig:nT.T15}}

\plotone{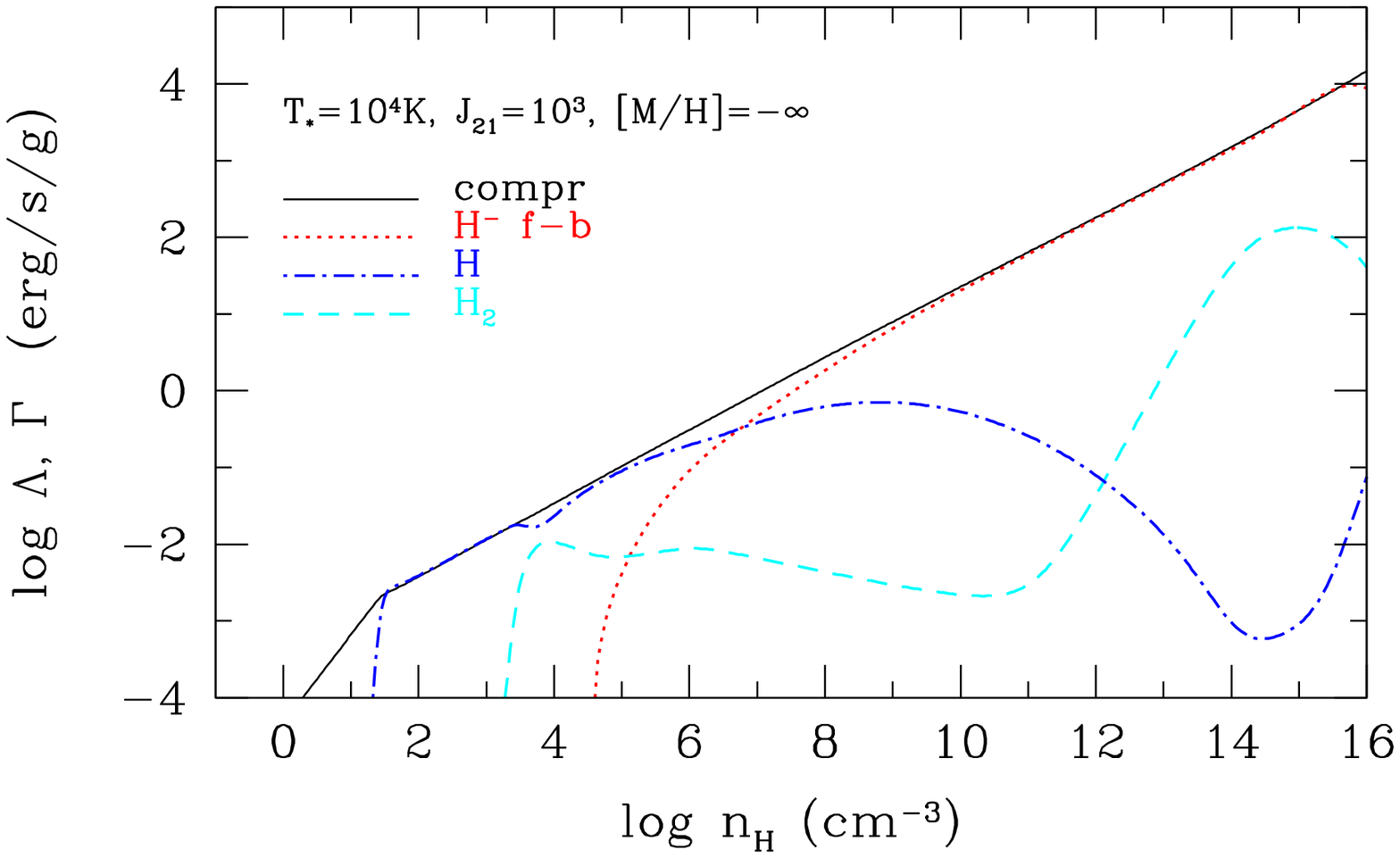} \figcaption[]{Contributions of various
processes to the total cooling rate, as a function of the number
density, for a metal-free cloud irradiated by an extremely intense FUV
field ($T_{\ast}=10^{4}$K, $J_{21}=10^{3}$; with the temperature
evolution shown in Figure \ref{fig:nT.T14}).
The meaning of the symbols is as follows: ``compr'' indicates 
compressional heating; 
``H$^{-}$ f-b'' cooling by  H$^{-}$ free-bound emission; 
``H''  cooling by  H line emission;
``H$_{2}$''  cooling by H$_2$ line emission. 
\label{fig:cool.T14J13D00M00}}

\plotone{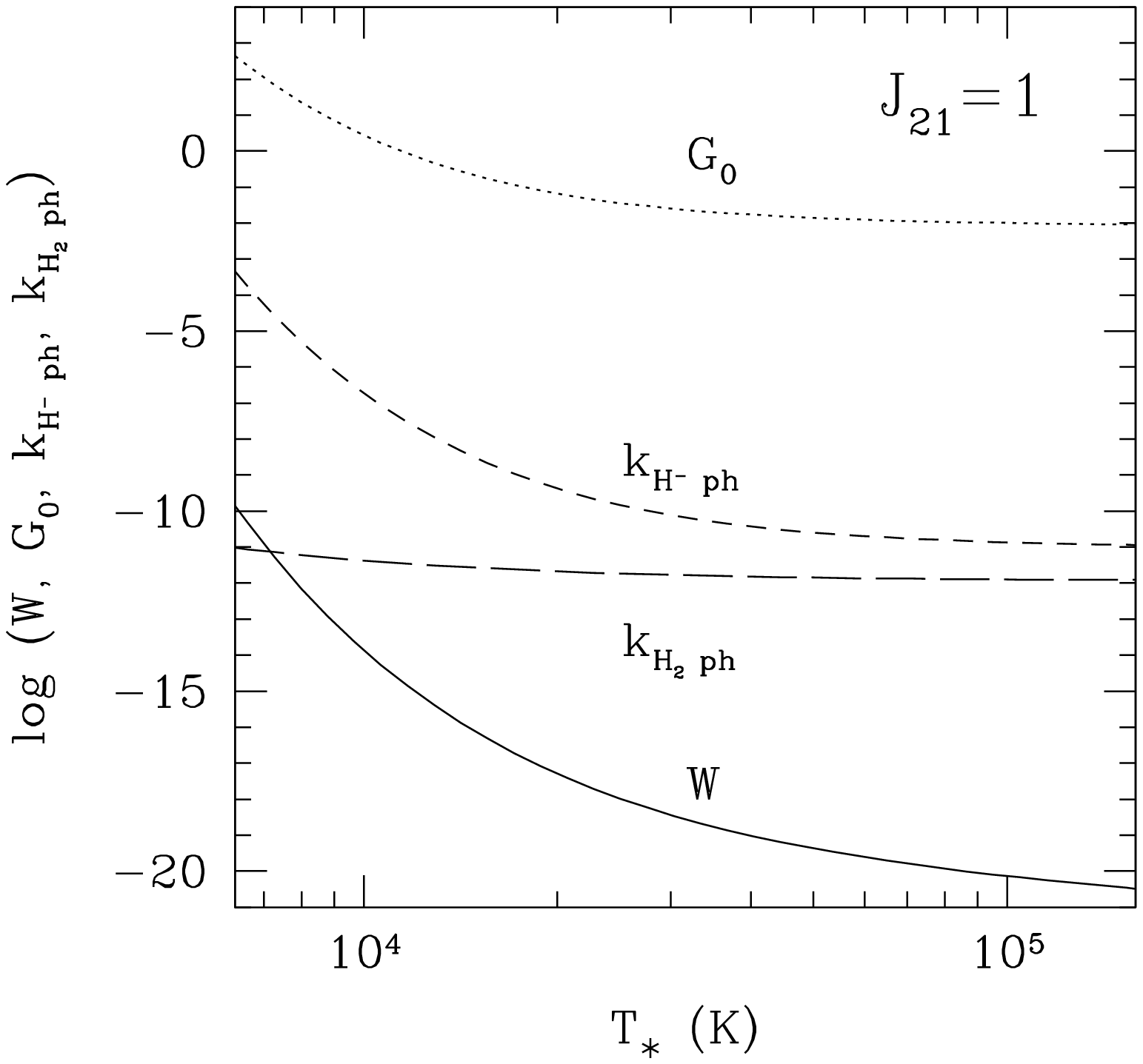} \figcaption[radparam]{H$_2$ and H$^-$
photodissociation rate coefficients as a function of radiation
temperature for $J_{21}=1$.  Also shown for reference are the Habing
parameter $G_{0}$ and dilution factor $W$.  
For a fixed intensity $J_{21}=1$ at 13.6~eV, the H$_2$ photodissociation
coefficient is almost constant with $T_{\ast}$ while the H$^{-}$
coefficient is about four orders of magnitude higher for
$T_{\ast}=10^{4}{\rm K}$, and thus reduces H$_2$ formation by a
large factor (see text).
\label{fig:radparam}}

\plotone{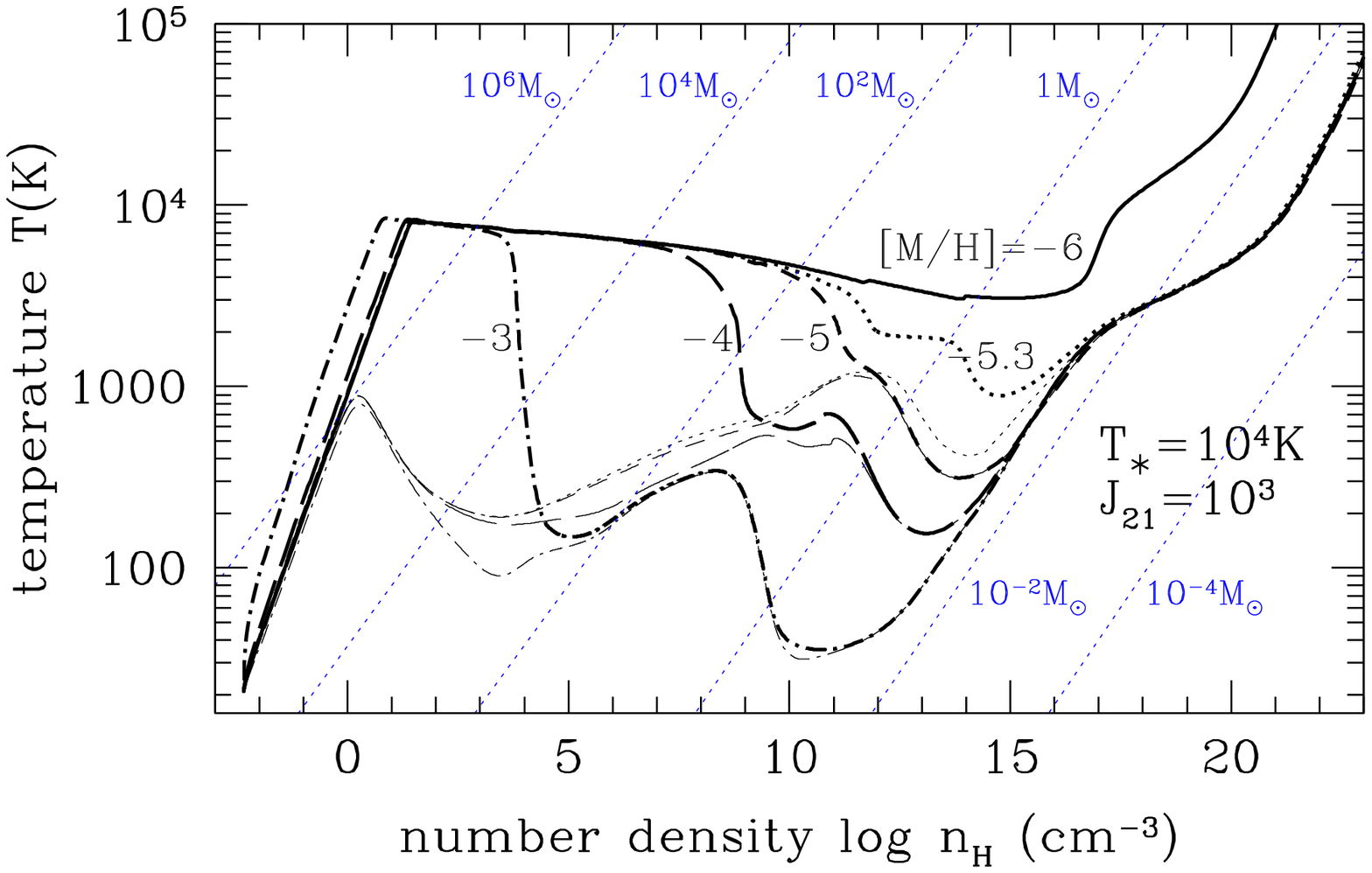}
\plotone{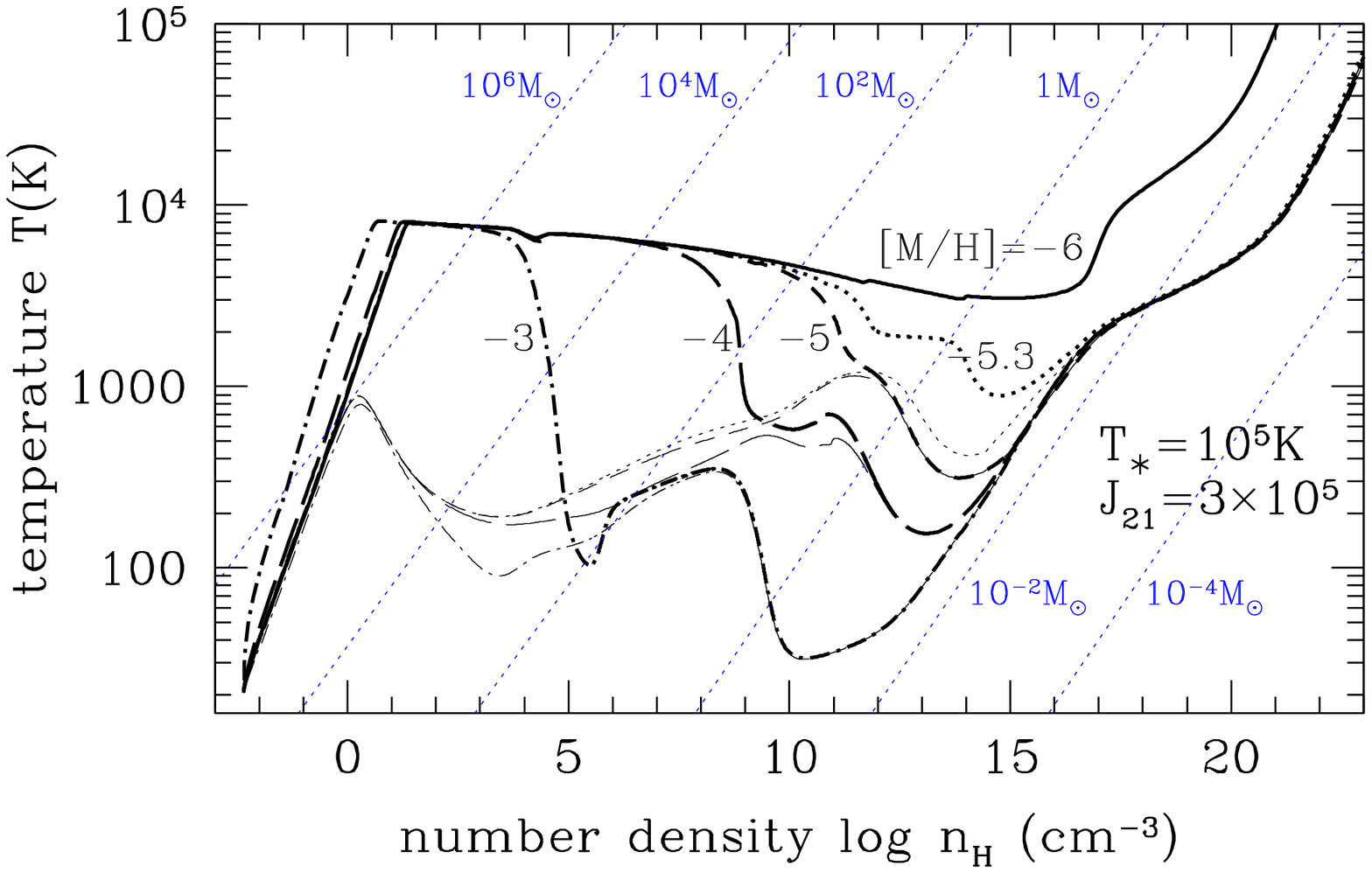} \figcaption[] {The temperature evolution of
clouds with initial metallicity [M/H]= -6 (solid), -5.3 (dotted), 
-5 (short-dashed), -4 (long-dashed), and -3 (dash-dotted)
irradiated by a FUV field with (a) $T_{\ast}=10^{4}$K, and
$J_{21}=10^{3}$, and (b)
$T_{\ast}=10^{5}$K, and $J_{21}=3\times10^{5}$.  Thin curves show the
results obtained without an external FUV field for initial
metallicities [M/H]= -5.3, -5, -4, and -3 (the same line types as 
the irradiated cases with the same metallicity).  
Due to photoelectric
heating, when [M/H]=-3, the temperature at the lowest densities is
higher than in the other models.
\label{fig:nT.JZ}}

\plotone{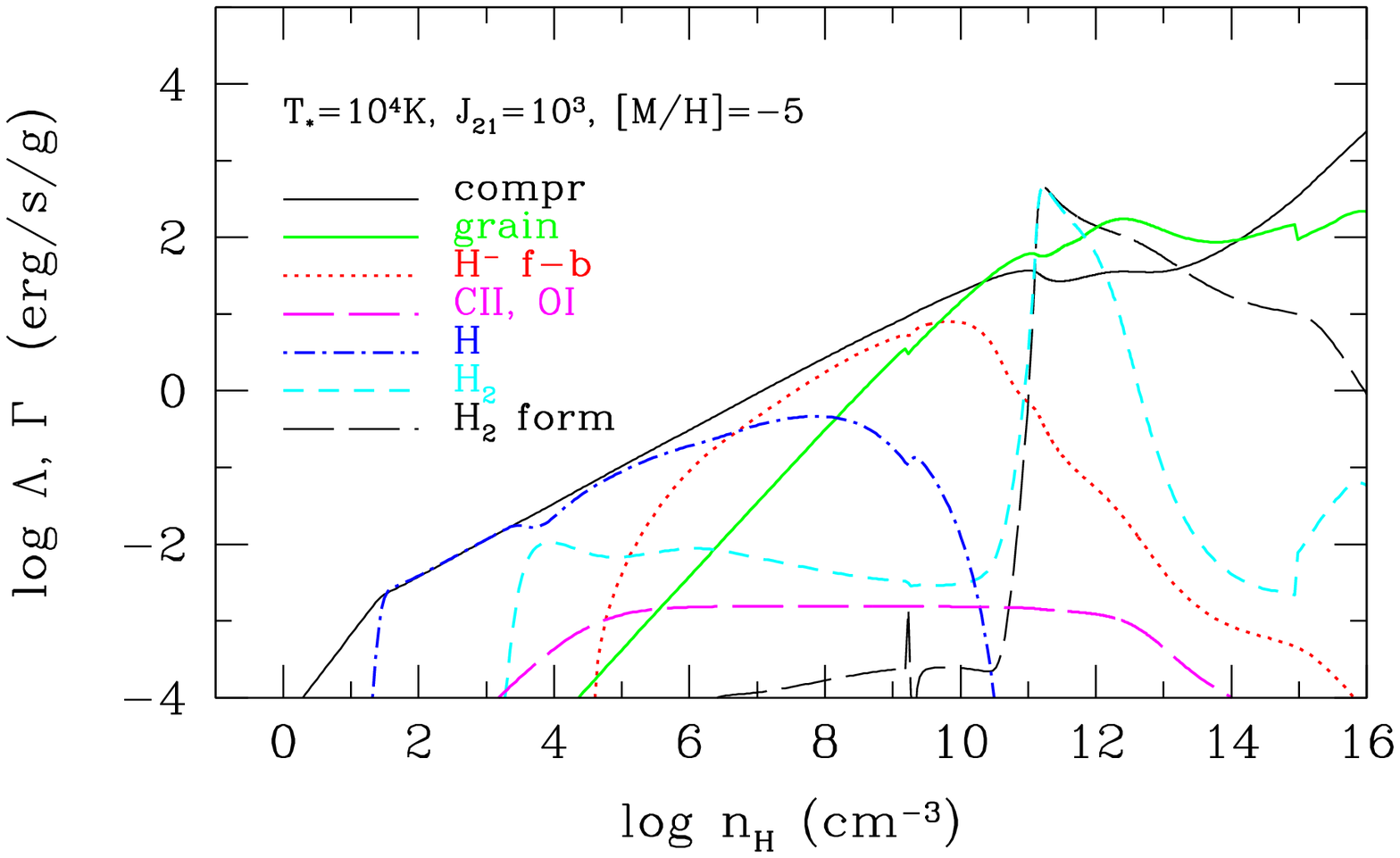} \figcaption[]{Cooling and heating
rates contributed by each process during the collapse of a cloud with
a metallicity of [M/H]=-5 and external FUV radiation with
($T_{\ast}=10^{4}$K, $J_{21}=10^{3}$).  
The corresponding temperature drop at $n_{\rm H} \sim 10^{10}{\rm cm^{-3}}$ 
(shown in Figure \ref{fig:nT.JZ}) is caused by dust cooling.
The meaning of the symbols is as follows: 
``compr'' indicates compressional heating; 
``grain''  cooling by dust thermal emission;
``H$^{-}$ f-b''  cooling by the H$^{-}$ free-bound emission; 
``CII, OI''  cooling by the CII and OI fine-structure line emission; 
``H''  cooling by the H line emission;
``H$_{2}$''  cooling by the H$_2$ line emission; 
``H$_{2}$ form''  heating by the H$_2$ formation.
\label{fig:cool.2}}

\plotone{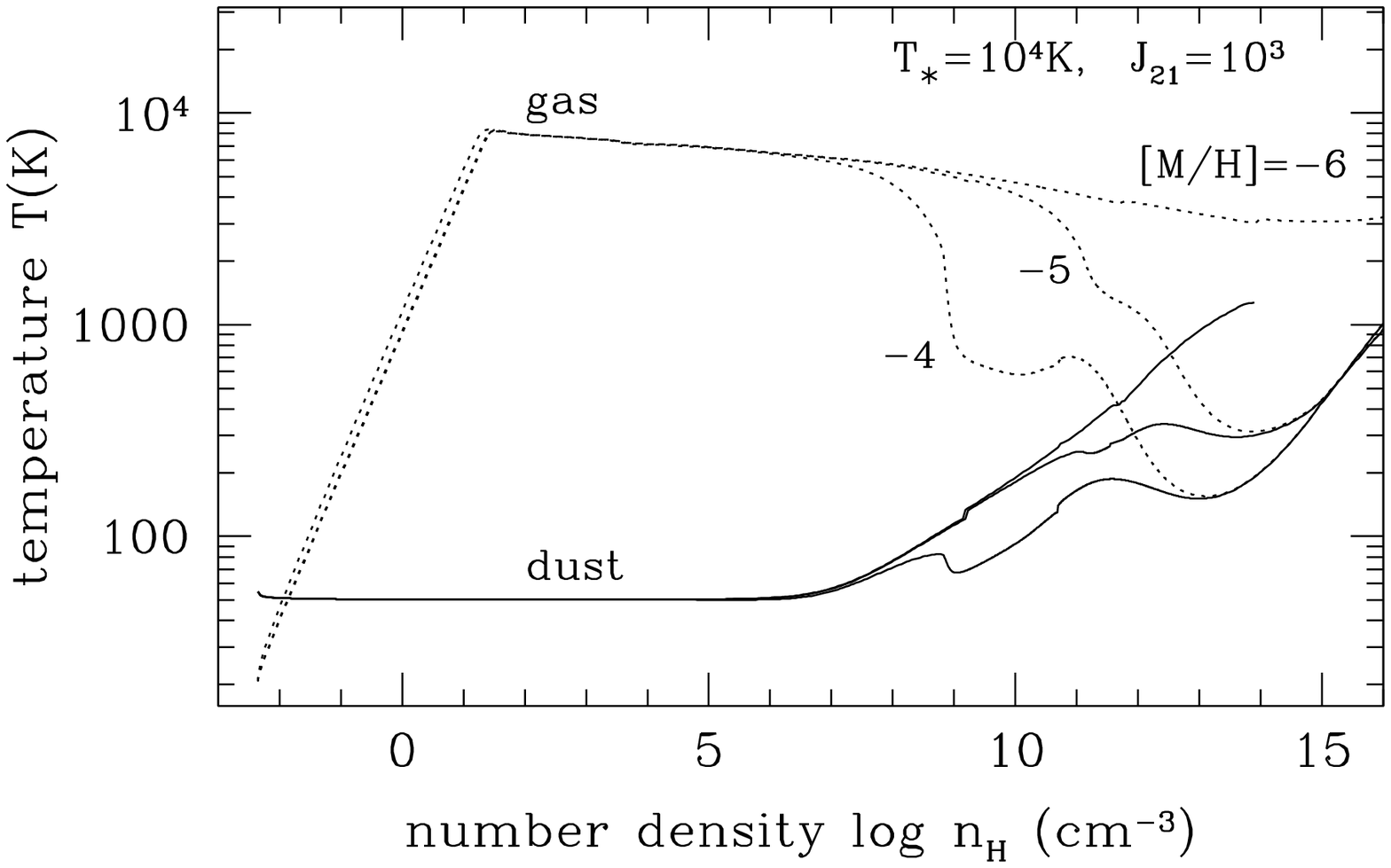} \figcaption[]{The dust temperature of clouds
(solid curves) with metallicity [M/H]=-6, -5, and -4 and with an
external radiation of ($T_{\ast}=10^{4}$K, $J_{21}=10^{4}$) as a
function of density. For the same models the dotted lines indicate the
corresponding gas temperatures.  The dust
temperature curve disappears at about $n_{\rm H} \sim 10^{14} {\rm cm^{-3}}$ 
for [M/H]=-6, reflecting the vaporization of the grains. 
\label{fig:nT.Tgr}}

\plotone{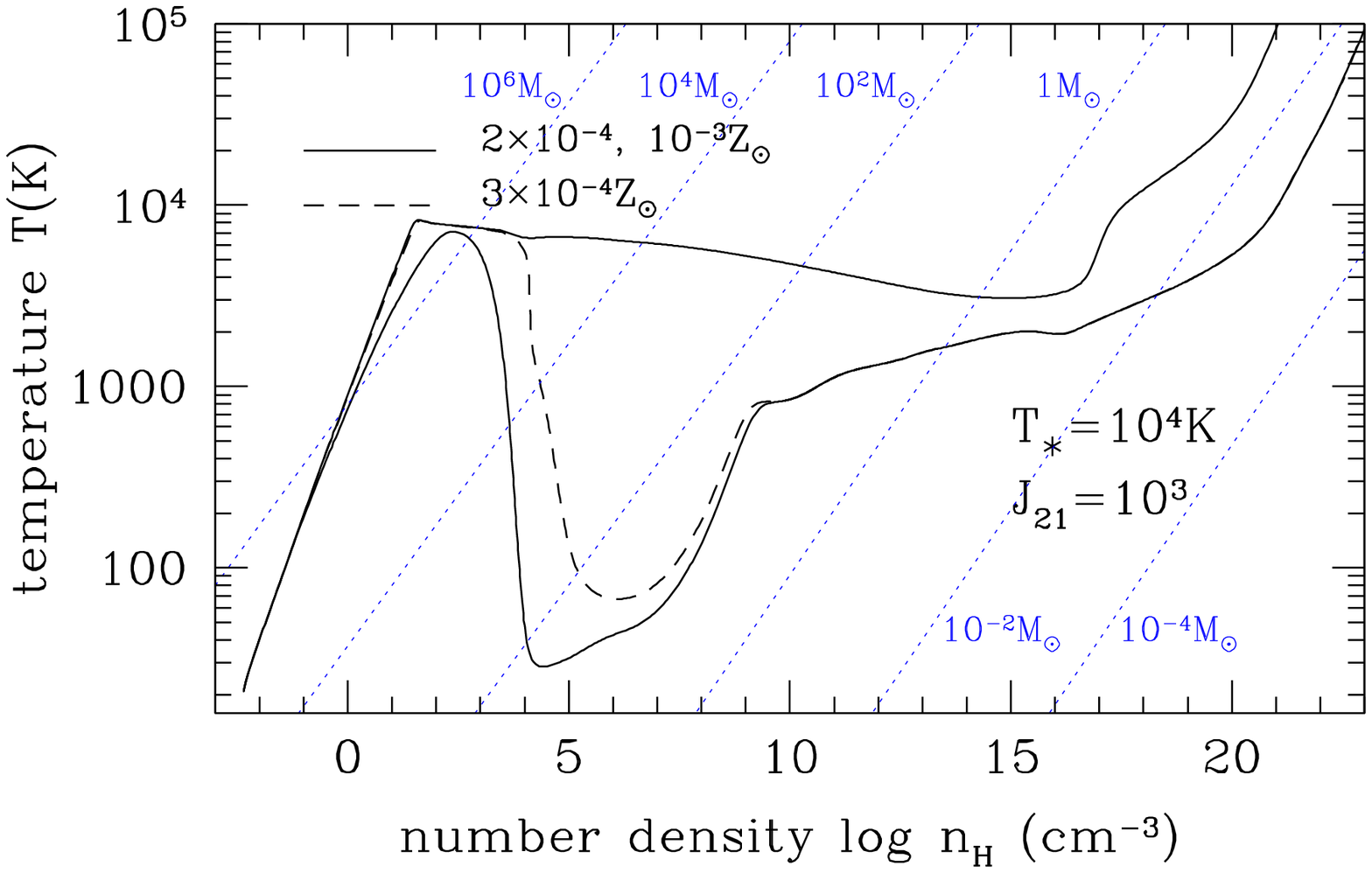} \figcaption[]{Temperature evolution of 
clouds when all metals are assumed to be in the gas phase (i.e. no
dust).  The radiation parameters are $T_{\ast}=10^{4}$K,
$J_{21}=10^{3}$.  Models with initial metallicities $Z=2 \times
10^{-4}Z_{\sun}$ (solid), $3 \times 10^{-4}Z_{\sun}$ (dashed), and
$10^{-3}Z_{\sun}$ (solid) are shown (from top to bottom).
\label{fig:nT.gasZ}}

\plotone{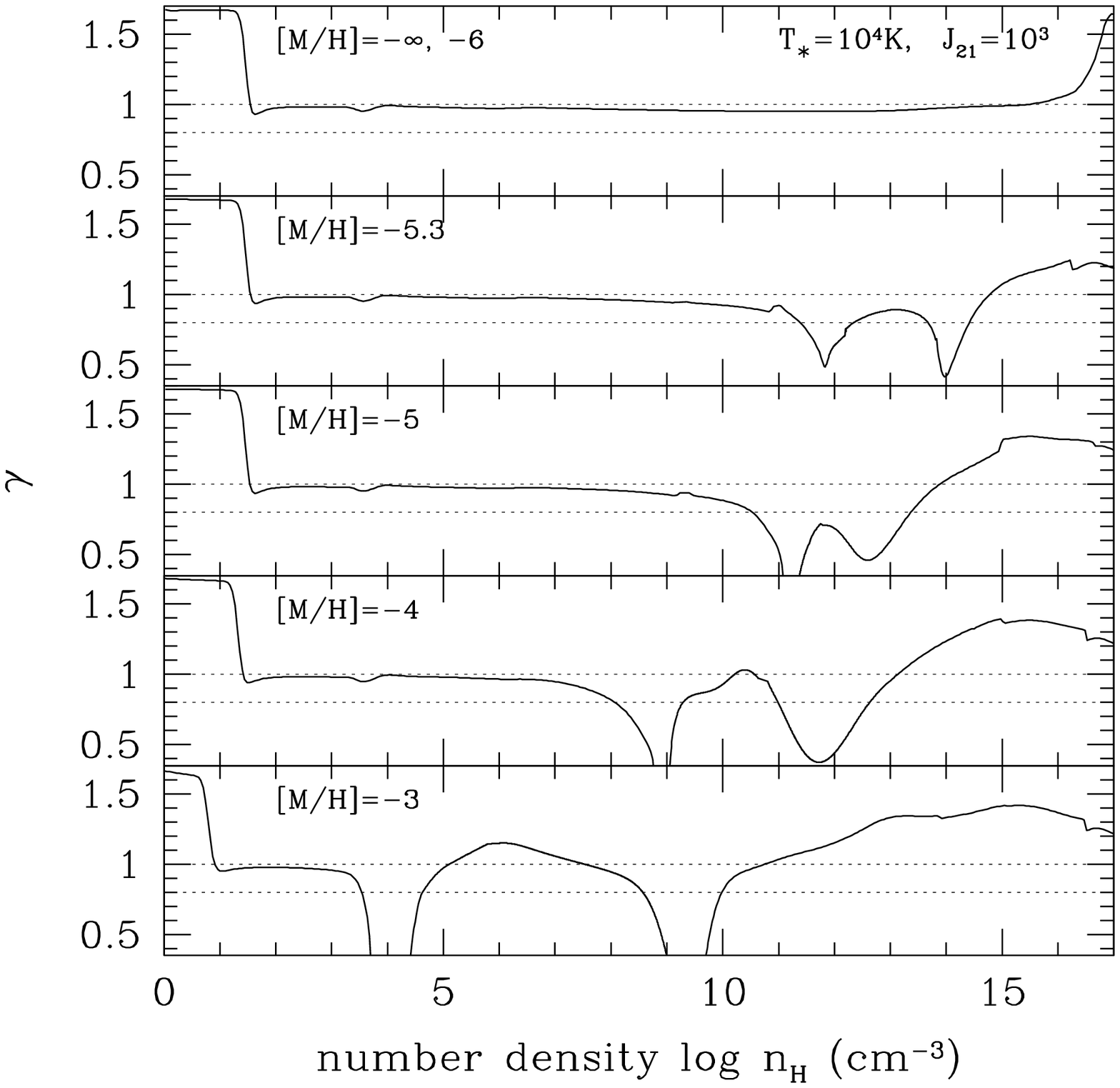} \figcaption[]{Evolution of effective adiabatic indices 
$\gamma = \partial {\rm ln}p/\partial {\rm ln}\rho$ for the models
shown in Figure \ref{fig:nT.JZ}(a), i.e., clouds 
irradiated by a field with parameters $T_{\ast}=10^{4}$K and $J_{\ast}=10^{3}$,
with metallicities [M/H]$=-\infty$, -6, -5.3, -5, -4, and -3 
(from top to bottom). 
The cases of [M/H]$=-\infty$ and -6 are identical and 
shown in the same panel (top).
The horizontal lines indicate $\gamma=0.8$ and 1.
The former is adopted as the fiducial value below which fragmentation 
is triggered. 
We assume that fragmentation stops when $\gamma$ becomes $\ga 1$.
\label{fig:gamma}}

\plotone{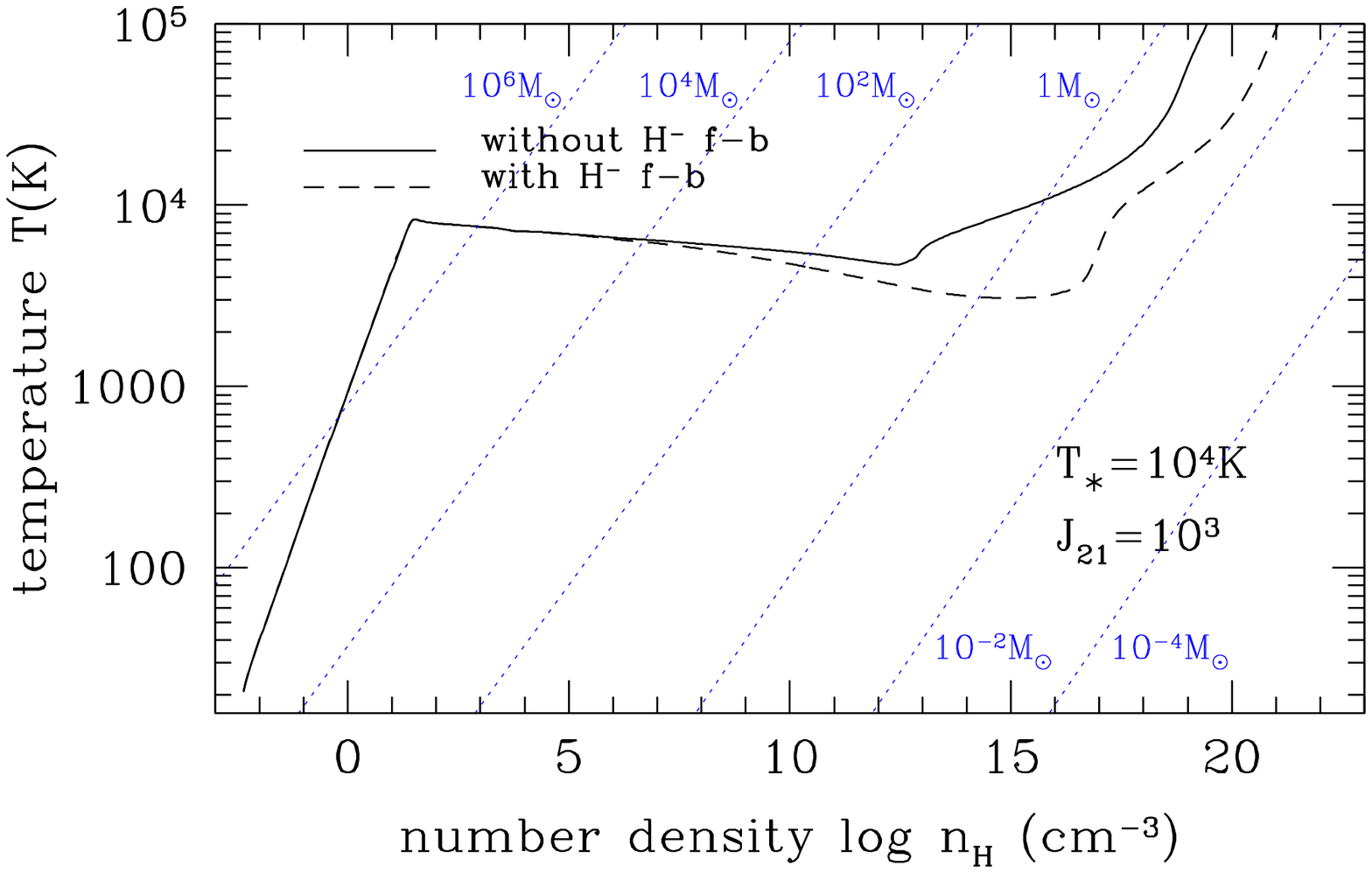} \figcaption[]{Effects of the H$^{-}$ free-bound
cooling on the temperature evolution.  The evolution of a metal-free
gas irradiated by a FUV field with ($T_{\ast}=10^{4}$K,
$J_{21}=10^{3}$) is shown with and without the H$^{-}$ free-bound
cooling.
\label{fig:Hmbf}}

\end{document}